\documentclass[aps,prd,twocolumn,nofootinbib,superscriptaddress,amsmath,amssymb,floatfix,10pt]{revtex4-1}
\usepackage{placeins}
\usepackage{bm}

\usepackage{graphicx}
\usepackage[font=small,labelfont=bf,
   justification=Justified,
   format=plain]{caption}

\usepackage{subcaption}

\usepackage[hmargin=1cm]{geometry}

\usepackage{mathtools}
\usepackage{graphics}
\usepackage{graphicx}
\graphicspath{{./images/}}
\usepackage{dcolumn}
\usepackage{bm}
\usepackage{dsfont}
\usepackage{amsmath,amssymb}
\usepackage{hyperref}
\usepackage{tabularx}
\usepackage{epstopdf}
\usepackage[normalem]{ulem}
\usepackage[usenames]{color}
\usepackage{multirow}
\usepackage{makecell}
\usepackage{float}
\usepackage{tabularray}
\usepackage{diagbox}
\usepackage[abs]{overpic}
\allowdisplaybreaks
\hypersetup{
    colorlinks=true,
    linkcolor=blue,
    filecolor=magenta,
    urlcolor=blue,
    citecolor=blue
}
\definecolor{red(ncs)}{rgb}{0.77, 0.01, 0.2}

\usepackage{array}

\newcolumntype{C}[1]{>{\centering\arraybackslash}m{#1}}

\def\emri#1{Extreme Mass-Ratio Inspiral#1 (EMRI#1)\gdef\emri{EMRI}}
\def\imbh#1{Intermediate Mass Black Hole#1 (IMBH#1)\gdef\imbh{IMBH}}
\def\smbh#1{supermassive black hole#1(SMBH#1)\gdef\smbh{SMBH}}
\def\bbh#1{binary black hole#1 (BBH#1)\gdef\bbh{BBH}}
\def\imbhb#1{intermediate mass black hole binary#1 (IMBHB#1)\gdef\imbhb{IMBHB}}
\def\hmns#1{hypermassive neutron star#1 (HMNS#1)\gdef\hmns{HMNS}}
\def\bh#1{black hole#1 (BH#1)\gdef\bh{BH}}
\def\ns#1{neutron star#1 (NS#1)\gdef\ns{NS}}
\def\hmns#1{hyper-massive neutron star#1 (HMNS#1)\gdef\hmns{HMNS}}
\def\bhns#1{black hole-neutron star#1 (BHNS#1)\gdef\bhns{BHNS}}
\def\nsbh#1{neutron star-black hole#1 (NSBH#1)\gdef\bhns{NSBH}}
\def\bns#1{binary neutron star#1 (BNS#1)\gdef\bns{BNS}}
\def\gw#1{gravitational wave#1 (GW#1)\gdef\gw{GW}}
\def\pnw#1{post-Newtonian#1 (PN#1)\gdef\pnw{PN}}
\def\eos#1{equation of state#1 (EOS#1)\gdef\eos{EOS}}
\def\gpu#1{graphics processing unit#1 (GPU#1)\gdef\gpu{GPU}}
\def\gr#1{General Relativity#1 (GR#1)\gdef\gr{GR}}
\def\cbc#1{compact binary coalescence#1 (CBC#1)\gdef\cbc{CBC}}
\def\nr#1{Numerical Relativity#1 (NR#1)\gdef\nr{NR}}
\def\hom#1{Higher Order Mode#1 (HOM#1)\gdef\hom{HOM}}
\def\qnm#1{Quasi Normal Mode#1 (QNM#1)\gdef\qnm{QNM}}
\def\emw#1{Electromagnetic Waves#1 (EMW#1)\gdef\emw{EMW}}
\def\snr#1{Signal to Noise Ratio#1 (SNR#1)\gdef\snr{SNR}}
\def\gr#1{General Relativity#1 (GR#1)\gdef\gr{GR}}
\def\psd#1{Power Spectral Density#1 (PSD#1)\gdef\psd{PSD}}
\def\asd#1{Amplitude Spectral Density#1 (ASD#1)\gdef\asd{ASD}}
\def\grb#1{gamma-ray burst#1 (GRB#1)\gdef\grb{GRB}}
\def\far#1{False Alarm Rate#1 (FAR#1)\gdef\far{FAR}}
\def\pe#1{Parameter Estimation#1 (PE#1)\gdef\pe{parameter estimation}}
\def\mcmc#1{Makov Chain Monte Carlo#1 (MCMC#1)\gdef\mcmc{MCMC}}
\def\rjmcmc#1{reversible-jump Markov chain Monte Carlo#1 (RJMCMC#1)\gdef\rjmcmc{RJMCMC}}
\def\mlw#1{maximum likelihood {\tt LALInference} waveform#1 (MLW#1)\gdef\mlw{MLW}}
\def\mbw#1{median {\tt BayesWave} waveform#1 (MBW#1)\gdef\mbw{MBW}}
\def\llo#1{LIGO Livingston Observatory#1 (LLO#1)\gdef\llo{LLO}}
\def\lho#1{LIGO Hanford Observatory#1 (LHO#1)\gdef\lho{LHO}}

\def\bw{{\tt BayesWave}}

\def\bl{{\tt BayesLine}}
\def\splusg{{\tt Joint}}
\def\cbcglitch{{\tt CBC$+$Glitch}}

\def\preml{$\mathrm{s}_{\mathrm{pre}}$}
\def\postml{$\mathrm{s}_{\mathrm{post}}$}
\def\gsplusg{$\mathrm{g}_{\mathrm{rec, J}}$}
\def\ggo{$\mathrm{g}_{\mathrm{rec, G}}$}

\newcommand{\CIT}{\affiliation{Department of Physics, California Institute of Technology, Pasadena, California 91125, USA}}

\newcommand{\LIGOCIT}{\affiliation{LIGO Laboratory, California Institute of Technology, Pasadena, California 91125, USA}}

\newcommand{\GA}{\affiliation{School of Physics, Georgia Institute of Technology, Atlanta, Georgia 30332, USA}}

\newcommand{\NASA}{\affiliation{NASA Marshall Space Flight Center, Huntsville, AL 35812, USA}}

\newcommand{\MSU}{\affiliation{eXtreme Gravity Institute, Department of Physics, Montana State University, Bozeman, Montana 59717, USA}}

\usepackage{xcolor}
\usepackage{comment}
\usepackage[normalem]{ulem}
\usepackage{soul}

\begin{document}

\title{Assessing and Mitigating the Impact of Glitches on Gravitational-Wave Parameter Estimation: a Model Agnostic Approach}

\author{Sudarshan Ghonge}
\GA

\author{Joshua Brandt}
\GA

\author{J. M. Sullivan}
\GA

\author{Margaret Millhouse}
\GA

\author{Katerina Chatziioannou}
\CIT \LIGOCIT

\author{James A. Clark}
\GA \CIT \LIGOCIT

\author{Tyson Littenberg}
\NASA

\author{Neil Cornish}
\MSU

\author{Sophie Hourihane}
\CIT

\author{Laura Cadonati}
\GA

\date{\today}

\begin{abstract}

In this paper we investigate the impact of transient noise artifacts, or {\it glitches}, on gravitational-wave inference from ground-based interferometer data, and test how modeling and subtracting these glitches affects the inferred parameters. Due to their time-frequency morphology, broadband glitches cause moderate to significant biasing of posterior distributions away from true values. In contrast, narrowband glitches induce negligible biasing effects, due to distinct signal and glitch morphologies. We inject simulated binary black hole signals into data containing three occurring glitch types from past LIGO-Virgo observing runs, and reconstruct both signal and glitch waveforms using \bw{}, a wavelet-based Bayesian analysis. We apply the standard LIGO-Virgo-KAGRA deglitching procedure to the detector data, which consists of  subtracting from  calibrated LIGO data the glitch waveform estimated by the joint \bw{} inference. {We produce posterior distributions on the parameters of the injected signal before and after subtracting the glitch,} and we {show that removing the transient noise} effectively mitigates bias from broadband glitches. This study provides a baseline validation of existing techniques, while demonstrating waveform reconstruction improvements to the Bayesian algorithm for robust astrophysical characterization in glitch-prone detector data.

\end{abstract}

\maketitle

\section{Introduction}
\label{sec:intro}

The Laser Interferometer Gravitational-wave Observatory (LIGO) \cite{aLIGO}, Virgo \cite{aVirgo} and KAGRA \cite{KAGRA}, form an international detector network for \gw{s}, 
 tiny ripples in spacetime produced by extreme astrophysical events. 
Since the first direct detection of merging black holes in 2015~\cite{Abbott_2016}, GWs are providing insights into the astrophysics of compact objects and the cosmology of the universe where they reside~\cite{GWTC3, GWTC3pop}. 
\gw{s} are detected by measuring changes in the relative length of km-scale arms in ground-based interferometers; these changes are infinitesimally small, 
of the order of $10^{-18}$ m  absolute length change, corresponding to a fractional length change, or \textit{strain}, of approximately $10^{-21}$.
The detection sensitivity is typically limited by seismic, thermal, and quantum noise effects.
In this paper, we focus on transient noise artifacts (\textit{glitches}) 
that could be confused with \gw{} signals. 

When correlated with auxiliary monitoring channels, some glitches can be mitigated {\cite{Detchar}}. For instance, the rate of high-power glitches at the \lho{} was reduced from 0.82 to 0.18 per minute by stabilizing optic suspension components; the resulting false-positive alert rate dropped from 55\% to 21\% {\cite{GWTC3, Davis:2018yrz}}. 
{Some studies have recently employed machine learning methods to subtract broadband noise} {\cite{Macas:2023wiw, Udall:2022vkv}} {and transient noise} {\cite{Ashton:2022ztk}}.

Unfortunately, the majority of noise transients do not have a well-identified source, and no software or hardware studies have been able to reliably identify one for them.  
Between November 2019 and March 2020, glitches occurred once every 0.32 minutes at \lho{} and 1.17 minutes at the \llo{}, and 17 of the 90 \gw{} events in the first three Gravitational Wave Transient Catalogs coincided within one second of a glitch~\cite{GWTC3}.
These unmitigated noise transients are problematic for both searches and downstream analyses. For searches, glitches could possibly mimic a \gw{} signal leading to a false detection, or increase the false alarm rate of a true detection thereby lowering the confidence in that detection\cite{Detchar}. For downstream analyses, a glitch temporally coincident with a real signal can reduce the parameter estimation accuracy\cite{Pankow_2018,Powell2018,Driggers2019,Payne2022}. In the work, we focus on the effect of glitches on parameter estimation.
The procedure of subtracting a glitch from data near a gravitational wave was first used in the data surrounding the first binary neutron star merger observed by LIGO and Virgo, GW170817, where a loud glitch occurred at \llo{} one second before merger~\cite{TheLIGOScientific:2017qsa, Pankow_2018}. Since then, glitch mitigation has become a routine ingredient of LIGO-Virgo analyses~\cite{GWTC2, GWTC3}.

In this paper, we investigate the impact on the accuracy of \gw{} parameter estimation of three common glitch classes: {\em blip}, {\em scattering}, and {\em tomte}, and we examine the improvements gained by {\em deglitching} the detector data. 
We select up to three occurrences of each glitch class~\cite{Chatziioannou:2021ezd} in a single detector and add a simulated \gw{} signal near the glitch time. 
We then use \bw{} to separate the signal, glitch, and noise components in the data by simultaneously modeling the signal and glitch parts using wavelets as described in \cite{bayeswave_III}. By subtracting the glitch component and performing \pe{} on this deglitched data, we can characterize the improvements in parameter inference, as a function of the time of separation between glitch and signal. Works such as \cite{Chatziioannou:2021ezd} and \cite{Hourihane:2022doe} have already demonstrated the effectiveness of a similar deglitching procedure with the additional improvement of replacing the wavelet frame from the signal part of \bw{} model with a spin-aligned dominant mode {\tt IMRPhenomD }\cbc{} waveform \cite{PhenomDI}. This is called the \cbcglitch{} configuration and is currently being used for deglitching data in the ongoing LIGO-Virgo-KAGRA fourth observing run. The methods described in this paper use the \bw{} configuration where both the signal and glitch powers are modeled using wavelets, and we will refer to this configuration as the \splusg{} configuration for the remainder of this paper.
Sections \ref{sec:motivation} and ~\ref{sec:methods} describe the data analysis methods and tools used. Section~\ref{sec:glitches} details the three main types of glitches we model with BayesWave. Section \ref{sec:results} summarizes the key results and findings. Finally, Section~\ref{sec:conclusion} concludes the analysis, and points to future, related work. \section{Gravitational Wave Parameter Estimation}
\label{sec:motivation}

The time series interferometric data in LIGO $\bm{h}$ is generally assumed to comprise of stationary Gaussian noise, denoted as $\bm{n}$,  occasional transient non-Gaussian features from astrophysical signals, $\bm{s}$ and instrumental glitches $\bm{g}$ \cite{Davis:2022dnd}:
\begin{equation}
    \bm{h} = \bm{s}  + \bm{n}+ \bm{g}.
    \label{eq:datadecomp}
\end{equation}

\gw{} \pe{} gives the posterior probability distribution of the parameters in the model for the gravitational wave signal $\bm{s}$ given the observed data.  The posterior probability distribution of $n$ parameters $\vec{\theta} = (\theta_1, \theta_2, ..., \theta_n)$ given \gw{} data $\bm{h}$, $p(\vec{\theta}|\bm{h})$ can be written as 
\begin{equation}
    p(\vec{\theta} | \bm{h}) = \frac{p(\vec{\theta}) p(\bm{h} | \vec{\theta})}{p(\bm{h})},
    \label{eq:bayes_thm}
\end{equation}
Where $p(\vec{\theta})$ is the prior distribution on the parameters of the model, $p(\bm{h})$ is the evidence (or marginalized likelihood), and $p(\bm{h} | \vec{\theta})$ is the likelihood. The likelihood is the probability of observing data $\bm{h}$ given some parameters $\vec{\theta}$, and defined by the models for both signal and noise.

Most \gw{} parameter estimation algorithms 
assume the data are a Gaussian process given by the \psd{} $S_n(f)$, that is 
the standard deviation of the routine Gaussian noise $\bm{n}$ at frequency $f$, plus a \gw{} signal. 
This reduces Eq.~\ref{eq:datadecomp} to $\bm{h}=\bm{s}+\bm{n}$, and the likelihood function is of the form
\begin{equation}
    p(\bm{h}|\vec{\theta}) \propto \exp\left( - \frac{1}{2} \langle \bm{h} - \bm{s}(\vec{\theta}) |\bm{h} - \bm{s}(\vec{\theta}) \rangle \right),
    \label{eq:likelihood}
\end{equation} 
where  $\langle \bm{a}|\bm{b} \rangle$ is the network noise-weighted inner product:
\begin{equation}
\langle \mathbf{a}|\mathbf{b} \rangle \equiv 4 \Re \, \sum_{i}^{n} \int_{0}^{\infty} \frac{\widetilde{a}^i(f)\widetilde{b}^{i*}(f)}{S_n^i(f)}df,
\label{eq:inner_prod}
\end{equation}
and $\widetilde{a}^{i}(f)$ is the Fourier transform of time series $a^{i}$ in the $i^{\mathrm{th}}$ detector.
 LIGO-Virgo \pe{} models the \gw{} signal as  a semi-analytical template from \gr{} (called an \textit{approximant}), 
 $\bm{s'}(\vec{\theta})$, resulting from the coalescence of objects described by the astrophysical parameters $\vec{\theta}$ \cite{Taracchini:2013rva, Nagar:2015xqa, Pan:2013rra, Ajith:2007kx}. Algorithms such as Bilby \cite{Bilby} sample $p(\vec{\theta} | \bm{h})$ using stochastic sampling methods such as nested sampling \cite{dynesty} or \mcmc{} \cite{Christensen:2001cr}.

If the parameter estimation analysis window contains both a \gw{} signal and an instrument glitch in one (or more) of the detectors, the assumption of Gaussian residuals in Eq.~\ref{eq:likelihood} breaks, leading to potentially biased posterior distributions of the astrophysical parameters \cite{Pankow_2018, Driggers2019, Hourihane:2022doe, Davis:2022dnd, Macas:2022afm} or incorrect inferences of effects like precession~\cite{Payne2022}. This could have impacts on downstream analysis such as inferences on population models \cite{GWTC3pop}. In the next section, we discuss a method for modeling and removing the glitch $\bm{g}$ from the data to ensure robust parameter estimation.

\section{Methodology}
\label{sec:methods}

\subsection{Glitch Subtraction with \bw{}}
\label{subsec:BW}
The LIGO-Virgo collaboration has developed a technique to subtract glitches from detector data using \bw{} ~\cite{cornish2015bayeswave,bayeswave_III, BW_repo}, a Bayesian algorithm that models non-Gaussian transients in \gw{} detector data as a superposition of sine-Gaussian (Morlet-Gabor) wavelets.
An individual wavelet is fully described by the parameter vector $\vec{\lambda}=\{t_0,f_0, Q, A,\phi_0\}$ which gives the wavelet's location and shape in time-frequency space. Analytically, a wavelet takes the form
\begin{equation}
    \Psi(t;\vec{\lambda}) = A\exp{\frac{(t-t_0)^2}{\tau^2}}\cos(2\pi f_0(t-t_0)+\phi_0)
   \label{eq:wavelEqn}
\end{equation}
where $\tau = \frac{Q}{2\pi f_0}$. 
These five parameters -- as well as the number of wavelets, $N$ -- are marginalized over with a reversible-jump (or transdimensional) Markov chain Monte Carlo (RJMCMC) algorithm \cite{green1995reversible}. 

For GW detector data containing a transient non-Gaussian artifact, \bw{} uses two models: 1. The data contain a coherent astrophysical signal (``signal model'') and 2. the data contain an incoherent instrumental glitch (``glitch model''). The signal model uses a set of wavelets common to all the detectors in the network and a set of extrinsic parameters which are also sampled via \bw{'s} RJMCMC sampler. The extrinsic parameters are the sky location and polarization information of the source which allow to forward project the signal onto each detector. The glitch model assumes no correlation between detectors and reconstructs the data with an independent set of wavelets in each one.

\begin{figure} 
    \includegraphics[width=0.75\columnwidth,clip=true]{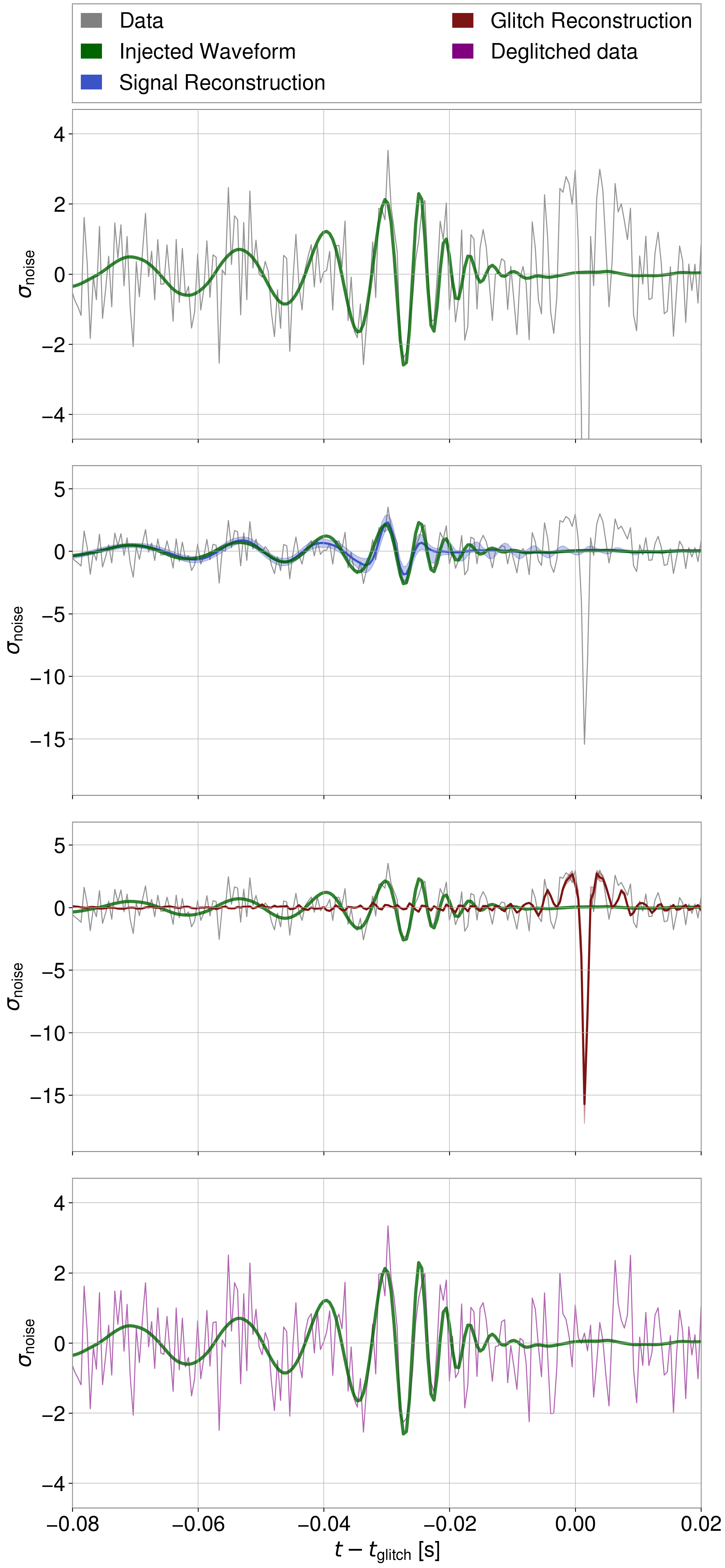}
    \caption[justification=Justified]{Time domain reconstructions for a sample BBH merger injection analyzed with the \splusg{} model. The injection's time of coalescence ($t_c$) precedes by 0.02 seconds the peak time of a blip glitch in \lho{} (at time 0 in this plot). {The panels in order from the top are: 1) Detector data (grey) with the added injection (green). 2) Median and 90\% credible intervals (blue) of the  Signal reconstruction from the \splusg{} model. 3) Median and 90\% credible intervals (red) of the glitch reconstruction from the \splusg{} model. 4) Deglitched detector data (magenta) with the added injection (green).}}
    \label{fig:joint_model_reconstruction}
\end{figure}

In the original implementation of \bw{}, the glitch and signal models are assumed to be disjoint (i.e. the data {contains} \textit{either} a signal \textit{or} a glitch), and the Bayes factor between the two models is used to measure the probability that the data contains a \gw{} signal~\cite{Kanner2016, Littenberg2016, O1AllSky, O2AllSky, O3AllSky}.
\bw{'s} \splusg{} now allows for both signal \textit{and} a glitch to simultaneously be present in the data~\cite{bayeswave_III}. { Figure \ref{fig:joint_model_reconstruction} shows the whitened strain for the detector data, the injected signal, and the reconstructions of the signal and glitch models. The whitened strain is obtained by filtering detector strain data with the inverse of the noise Amplitude Spectral Density (ASD). The ASD for the analysis window is calculated by \bl{} as part of \bw{} analysis simultaneously with the \splusg{} model \cite{bayesline_2015, Chatziioannou:2019zvs}.} 

By simultaneously modeling the coherent and incoherent power, we can subtract the glitch from the data:
\begin{equation}
    \bm{h}_{DG}=\bm{h}-\bm{g}_{rec}
    \label{eq:bwRes}
\end{equation}
where $\bm{h}_{DG}$ is the deglitched data to be used for parameter estimation. For $\bm{g}_{rec}$ we 
select a random sample from the posterior distribution of $\bm{g}$ produced by \bw{}.

\subsection{Updated prior ranges}
\label{subsec: priors}

Previous detector characterization studies have identified several glitch classes with distinctive time-frequency features~\cite{GSpy}. We use this information to introduce separate Bayesian prior ranges on the glitch part of the \splusg{} model to more effectively and efficiently separate the signal and glitch power. In this section, we introduce these new glitch-specific prior ranges.

\paragraph{$Q$ prior.} \par
A wavelet quality factor $Q$ is proportional to the number of cycles in a wavelet.
For a given central frequency $f_0$, the time duration of a wavelet will increase with increasing $Q$. For previous analyses which focus on stellar-mass \bbh{} waveforms the prior on $Q$ for both the signal and glitch models was uniform in the range $[0.1,40]$. This prior range is generally sufficient for \bbh{} signals which have a duration less than 1 second, but to reconstruct glitches with duration greater than 1 second, such as scattering glitches, a higher upper bound on $Q$ allows for a smaller number of longer-duration wavelets. 

\paragraph{Dimensionality prior.}\par
\bw{} uses a reversible-jump MCMC to sample a variable dimensional parameter space by adding or removing wavelets. In the RJMCMC, the number of wavelets $N$ is a parameter in the model, along with the wavelet parameters described in Sec.~\ref{subsec:BW}.
This transdimensional sampling procedure balances model simplicity with goodness-of-fit so that we can accurately reconstruct the non-Gaussian feature in the data without overfitting.  
The number of wavelets used on average to reconstruct a waveform depends on its time-frequency morphology, but typically features with higher SNR require more wavelets~\cite{Littenberg2016, Kanner2016, Millhouse:2018dgi}. 
The default prior on the number of wavelets, $p(N)$, described in Ref.~\cite{bayeswave_III} was determined empirically from data from LIGO's first observing run (O1). $p(N)$ peaks at $N=3$. For computational reasons, there is typically a maximum of $N=100$ set.  For short-duration \gw{} signals of moderate SNR (SNR$<$100), the posterior on $N$ is far less than this maximum. 
However, glitches can be significantly louder than \bbh{} signals and may require larger than 100 wavelets. 
In this study, we modify \bw{} to allow for distinct upper bounds on the number of wavelets for each model component of the \splusg{} model.

\paragraph{Time prior.}\par
The prior on the wavelet central time $t_0$, also known as the \textit{window}, does not span the full length of data used in the \bw{} analysis since \bbh{} signals typically encountered in LIGO and Virgo are less than a second long. For stellar-mass \bbh{} systems we use a minimum analysis segment duration of 4 seconds to ensure a fine enough frequency resolution, but set the prior on $t_0$ to a uniform distribution over 1 second centered around the expected time of coalescence of the signal. A smaller range on $t_0$ limits the volume of the parameter space and ensures that the wavelets are only placed at times where we most expect the signal. This is computationally advantageous as it results in faster convergence of the RJMCMC chain. Details of the computational costs can be found in the appendix section of \cite{bayeswave_III}.
By default, the time prior is shared between the signal and glitch parts of the \splusg{} model. However, to analyze cases where the characteristic durations of the \gw{} signal and glitch are different, we allow for independent priors on $t_0$ for the glitch and the signal. For example, scattering glitches typically have durations of 3-4 seconds which is larger than the typical 0.2-1 second duration of stellar mass \bbh{} signals.

The choices for these prior ranges and analysis windows are informed by spectrograms of the data near the event in question, as well as preliminary information from the search pipelines.  Previous reconstruction studies~\cite{Littenberg2016, Millhouse:2018dgi,Pannarale2019} have looked in detail at how the number of wavelets used by \bw{} scales with the SNR and total mass of the signal for compact binary coalescences.  Preliminary parameter estimates from the search pipelines can then be used to determine the priors for the signal model.  Sepctrograms of the data are used to identify the type of glitch and/or estimate its properties such as central time, frequency, duration, and bandwidth.  This information can be used to set the glitch model priors.

 \section{Overview of Glitch Types}
\label{sec:glitches}
In this study, we focused on three types of glitches (blip, scattering, and tomte) which are commonly seen in LIGO data. 
We use the same GPS times used in~\cite{Hourihane:2022doe} which were confidently classified as containing a glitch from the citizen science GravitySpy Project \cite{GSpy, Zevin:2023rmt}. We use one additional glitch of the ``tomte'' class not included in~\cite{Hourihane:2022doe}. The sample glitches used here are from both O2 and O3 and the data is taken from the Gravitational-Wave Open Science Center \cite{gwosc_o1_o2, gwosc_o3}.
While these sample glitches use data from the two LIGO detectors, these procedures can also be applied to Virgo data.
An overview of each glitch type is given in the following sections. 

\vspace{20px}
\subsection{Blip}
\begin{figure} [htbp]
    \includegraphics[width=0.9\columnwidth,clip=true]{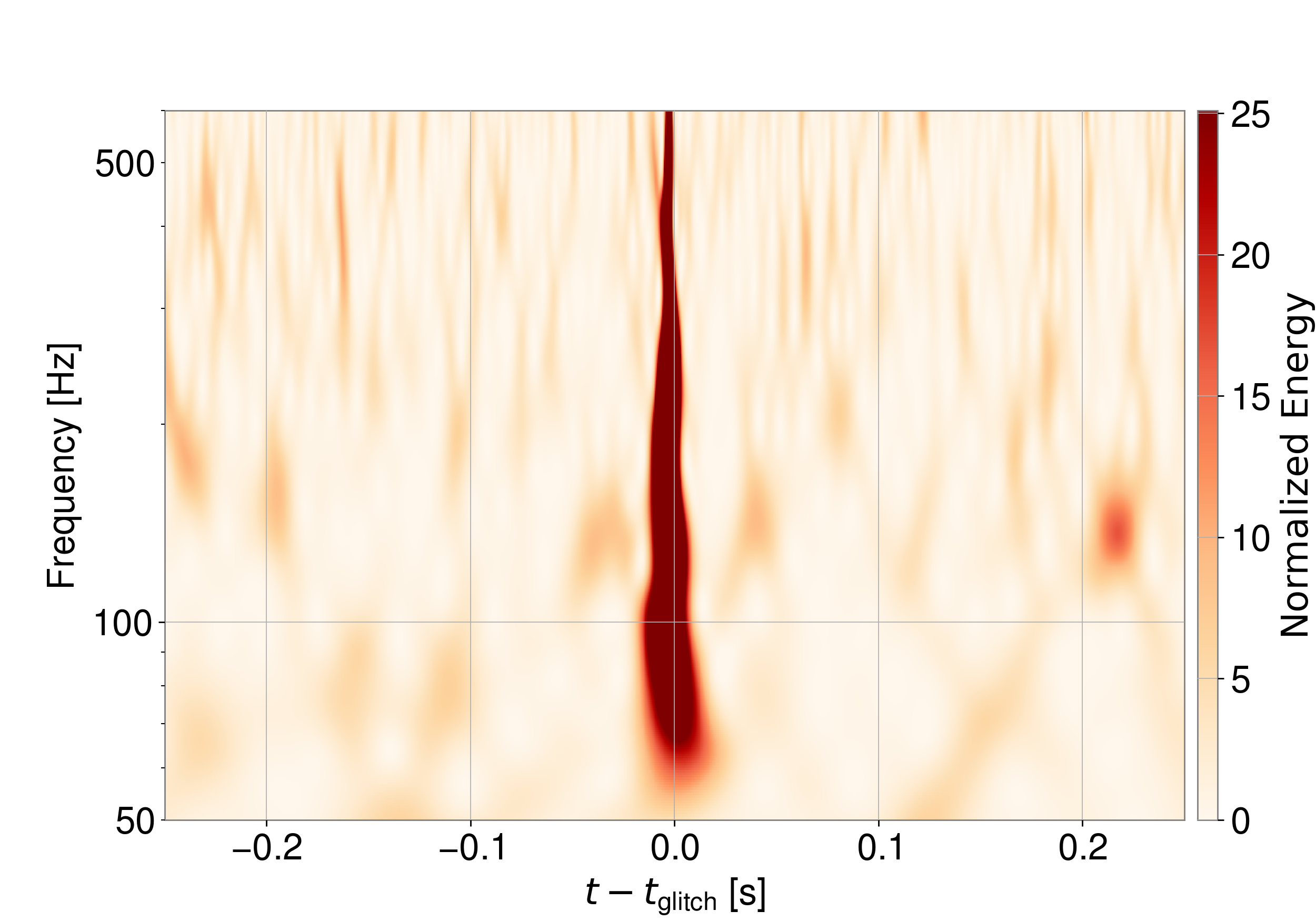}
    \caption{Time-frequency representation of \lho{} strain data containing blip glitch 1. The tear-drop-shaped morphology resembles the last few cycles of stellar-mass BBH inspirals}
    \label{fig:blip_spectrogram}
\end{figure}

{\em Blip} glitches are broadband, short-duration noise transients, with a typical duration of 10~ms and 20--1000 Hz frequency bandwidth. Figure \ref{fig:blip_spectrogram} shows the time-frequency (TF) representation of the \lho{} data containing blip glitch 1, depicting the absolute value of the detector data as functions of time and frequency. 
Blip glitches can contribute to the detection false-alarm-rate {(FAR)} 
due to their similarity with the final few cycles of stellar-mass binary coalescences ($20 M_{\odot} < \mathrm{M}_T < 150 M_{\odot}$ ) \cite{CaberoLundgren}, as well as to their  
relatively high incidence rates in both LIGO detectors: approximately every 15 minutes in \llo{} and every 30 minutes in \lho{} during O3.  No conclusive environmental or technical causes of blips have been identified to date~\cite{CaberoLundgren, Detchar}. 

\FloatBarrier
\subsection{Tomte}
\begin{figure} [htbp]
    \includegraphics[width=0.9\columnwidth,clip=true]{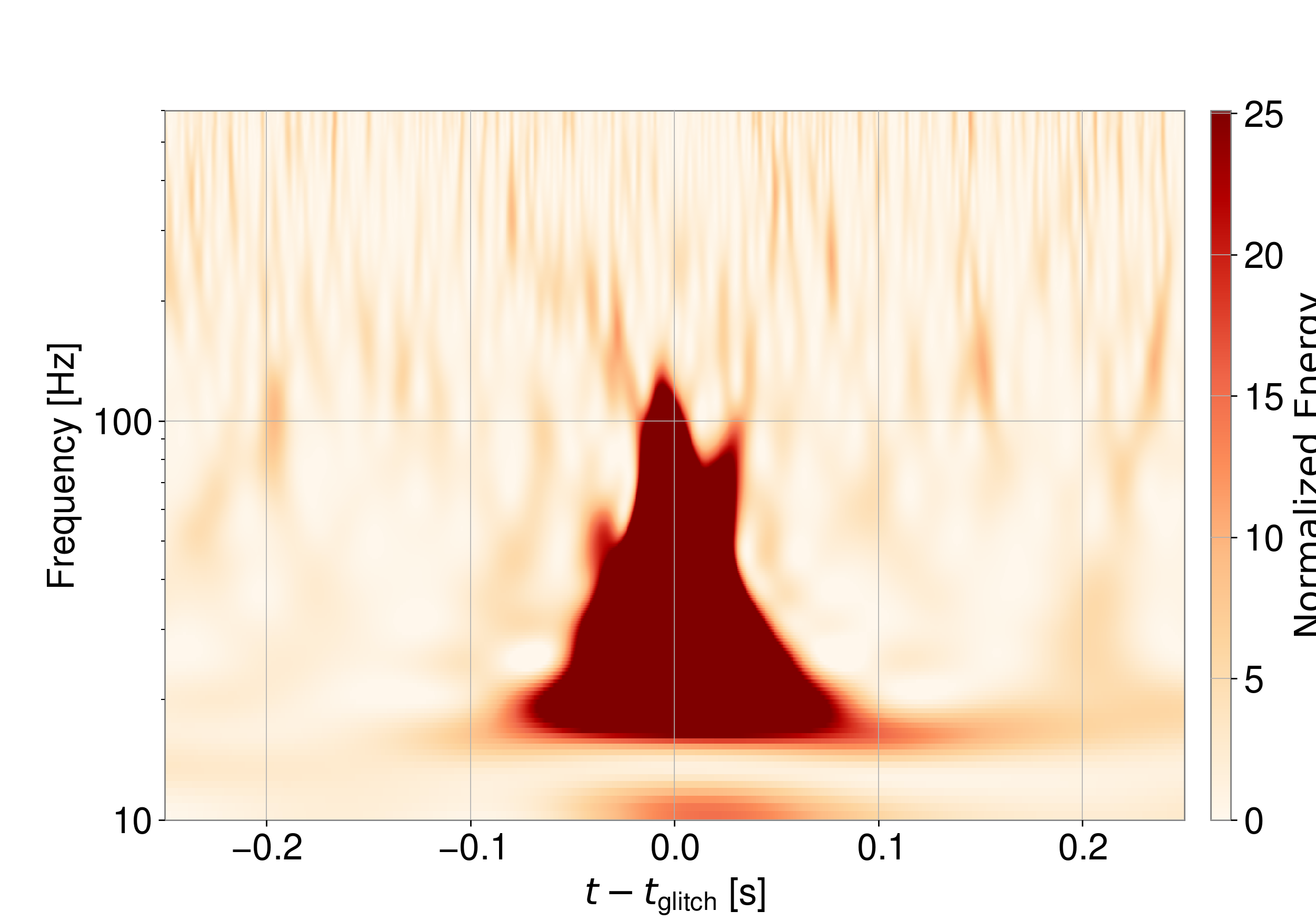}
    \caption{Tomte glitch 1 analog of Figure \ref{fig:blip_spectrogram}. Tomte glitches are so named due to the similarity in their appearance with the silhouette of the Nordic mythological creature of the same name. Tomtes are similar to blips but have smaller frequency spread (10 - 100 Hz) and longer durations ($\sim 100$ ms). Since the majority of their power lies at frequencies below $<100$ Hz, they are particularly problematic for high mass \bbh{} systems ($\mathrm{M}_T > 100 M_{\odot}$)}
    \label{fig:tomte_spectrogram}
\end{figure}

{\em Tomte} glitches owe their name to their appearance in a time-frequency map (see Figure \ref{fig:tomte_spectrogram}), which looks similar to the hat of the Nordic mythological creature of the same name.
During O3, they constituted 19\% of glitches in \lho{}, at the rate of  6 per hour. They were less frequent in \llo{}, where they occurred approximately once every 4 hours \cite{Glanzer_2023}.
Compared to blips, these glitches typically last longer ($\approx$100 ms), and peak frequency power in the 20--256 Hz bandwidth. They often better correspond to the frequency profile of high-mass binary signals than blips~\cite{Detchar}, and are especially problematic for how they bias parameter estimation. 
Like blips, however, no reliable determination of the source of tomte glitches has been made.

\FloatBarrier
\subsection{Scattering}
\begin{figure} [htb]
    \includegraphics[width=0.9\columnwidth,clip=true]{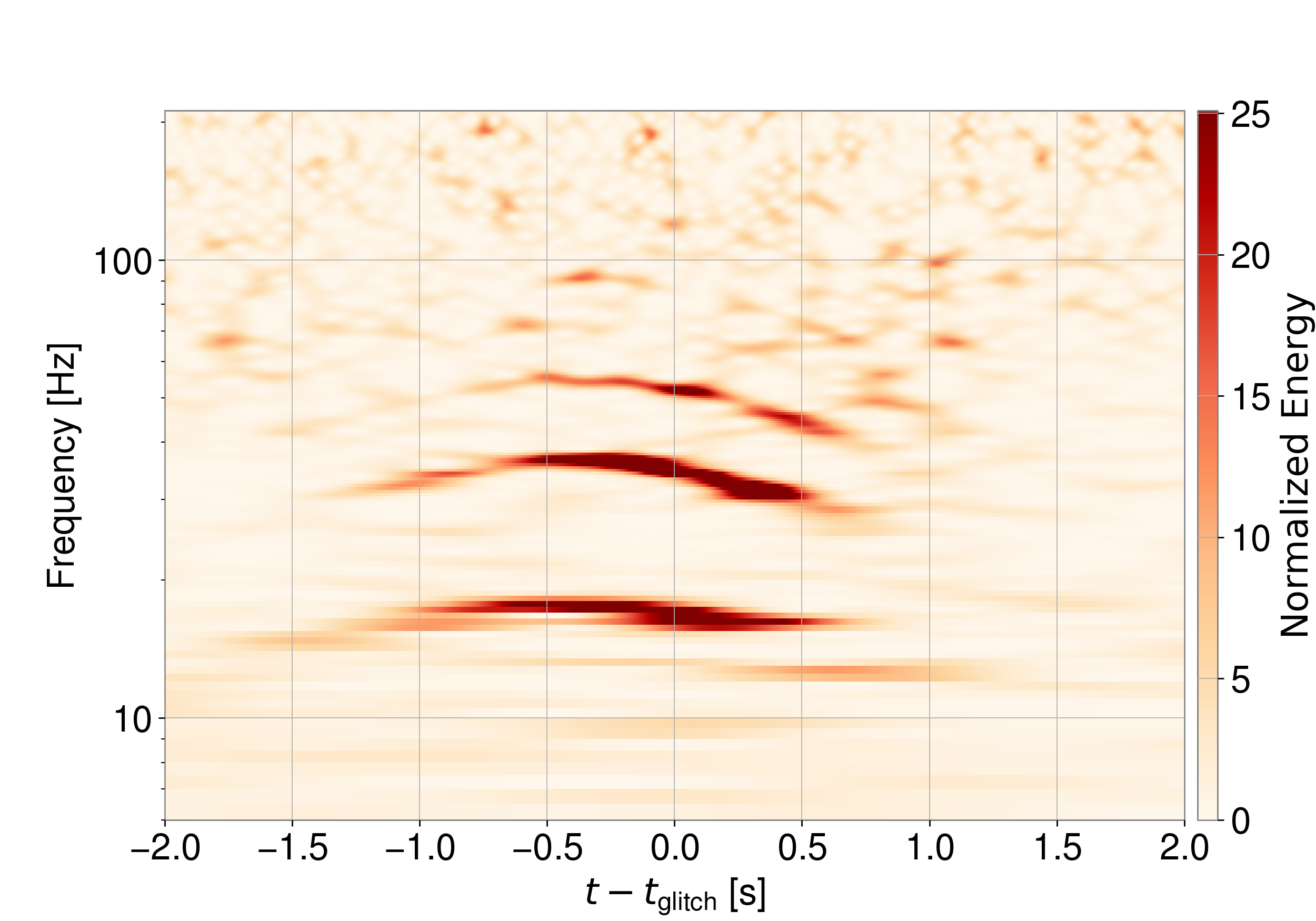}
    \caption{Scattering glitch 1 analog of Figure \ref{fig:blip_spectrogram}. Scattering glitches are of longer durations ($\sim 2$ s) and the power is spread into multiple harmonics. The primary frequency band is typically between 10 - 20 Hz and the highest harmonic is at frequencies $<100$ Hz. Due to the near-constant frequency of an individual harmonic, scattering glitches can resemble the inspiral part of stellar mass \bbh{} mergers.}
    \label{fig:scattering_spectrogram}
\end{figure}

{\em Scattering} glitches are characterized by their arching structure in the 10--120 Hz frequency band and   3–4 s duration~\cite{SoniBerry}, shown in Figure~\ref{fig:scattering_spectrogram}. Their ultimate causes are the $0.1 - 0.5$ Hz microseismic ground motion from waves crashing on the shores of nearby oceans \cite{SoniBerry}, and $0.03 - 0.1$ Hz earthquakes \cite{GWTC3}. The motion of the detector test masses leads to light scattering from and recombining back into the coherent beam in the form of scattering glitches' signature harmonic arches \cite{Detchar}. Scattering glitches are problematic due to the resemblance that their higher order harmonics can bear to inspiral signals of low-mass coalescing binaries ($ \mathrm{M}_T < 20 M_{\odot} $). During O3, scattered light glitches made up approximately $40\%$ of confidently classified glitches, occurring about 10 times per hour in both detectors \cite{Glanzer_2023}. Successful hardware mitigation efforts were carried out for scattering glitches during O3 \cite{redScattSoni}. Additionally, works such as \cite{Udall:2022vkv} introduced a physically motivated parametric model to reconstruct these glitches.

 \section{Analysis}
\label{sec:results}

To assess the impact of deglitching on parameter estimation, we add a simulated signal (referred to as an ``injection'') into real data containing the glitches described above.  The injected signal has source parameters similar to GW150914 \cite{PEofGW150914} and its luminosity distance scaled such that its \snr{} is equal to 15, as listed in Table \ref{table:injected_parameters}.
We use the Bilby software package \cite{Bilby} for template-based \pe{}, utilizing the  {\tt IMRPhenomPv2} approximant \cite{PhenomDI} to model $\bm{s}$, and the dynesty \cite{dynesty} nested-sampler to sample $p(\bm{h}|\vec{\theta})$ . 
We conduct \pe{} before ($\bm{h} = \bm{s} + \bm{g} + \bm{n}$) and after deglitching ($\bm{h}=\bm{s}+\bm{n} + \bm{g} - \bm{g}_{\textrm{rec}}$).

We inject the signal at various times around the glitch, and evaluate the change in PE before and after subtracting the glitch with \bw{},  and how this depends on the difference in the time of injected signal coalescence and the central time of the glitch as discussed in Section \ref{sec:glitches}.

 \begin{table}[ht]
\begin{center}
\begin{tabular}{lc c }
\hline\hline
Parameter & & Value \\
\hline
Mass 1 ($m_1$) & & $36 M_{\odot}$  \\
Mass 2 ($m_2)$ & & $29 M_{\odot}$  \\
Luminosity Distance ($d_L$) & & $1206$ Mpc \\
Right Ascension ($\alpha$) & & 6.14 \\
Declination ($\delta$) & & 0.81 \\
Zenith Angle ($\theta_{jn}$) & & 0 \\
Spin 1 ($a_1$) & & 0 \\
Spin 2 ($a_2$) & & 0 \\
$\phi_{12}$ & & 0 \\
$\phi_{jl}$ & & 0 \\
\hline\hline
\end{tabular}
\caption{{Detector-frame} source parameters for the simulated signal used in this study. This is the same injection as is used in \cite{Chatziioannou:2021ezd} and the luminosity distance is chosen such that the injection's signal-to-noise ratio (SNR) is equal to 15.}\label{table:injected_parameters}
\end{center}
\end{table}

We focus on three aspects:

\noindent 1. {\bf Effect of glitch on PE.} For each glitch class, we quantify the bias in the recovered posterior probability distributions of 
chirp mass, mass ratio, and $\chi_{\mathrm{eff}}$ as a function of the difference in the peak time of the glitch ($t_{\mathrm{glitch}}$) and the time of binary coalescence ($t_{\mathrm{signal}}$).
We also calculate the overlap between the maximum likelihood recovered (\preml) and injected waveform ($s$):
\begin{equation}
{\cal{O}} \equiv \frac{\langle \bm{a} | \bm{b} \rangle}{\sqrt{\langle \bm{a}|\bm{a} \rangle \langle \bm{b}|\bm{b} \rangle}}.
\label{eq:network_overlap}
\end{equation}
which ranges between $-1$ (complete disagreement), $0$ (no agreement), and $1$ (complete agreement).

\noindent 2. {\bf BW deglitching efficacy.} We show the overlap between the glitch reconstruction recovered (\gsplusg) by the \bw{} \splusg{} model on the data containing the injection ($\bm{h} = \bm{s} + \bm{g} + \bm{n}$) and  and the glitch reconstruction (\ggo) recovered by \bw{} on the data without an injection ($\bm{h} = \bm{g} + \bm{n}$).

\noindent 3. {\bf PE improvements after deglitching.} 
We compare the recovered posterior probability distributions of the above-mentioned parameters (chirp mass, mass ratio, and $\chi_{\mathrm{eff}}$) before and after deglitching. We also calculate the overlap between the maximum likelihood recovered (\preml) and injected waveform ($s$)

\subsection{Blip}
We analyze three blip glitches, two from \lho{} and one from \llo{} at GPS times 1168989748.13 (\textbf{20 Jan.
    2017}), 1165578732.45 (\textbf{12 Dec. 2016}), and 1171588981.76 (\textbf{19 Feb. 2017}) with \snr{s} of 20.1, 22, and 19 as computed by the Omicron pipeline \cite{Omicron}. For each glitch, we inject the signal at 9 times spaced 0.025 seconds around the glitch peak time. For each injection, we apply the above-described procedure. Since the signal and glitch, both have durations of less than 1 second and are morphologically similar, we use the default ranges for priors described in Section \ref{sec:methods} B, i.e., default values for the quality factor ([0.1, 40]), model dimensionality ([1, 100]), analysis segment (4 seconds) and window (1 second) for both the signal and glitch parts of the \splusg{} model.

\begin{figure}
    \includegraphics[width=\columnwidth, clip=true]{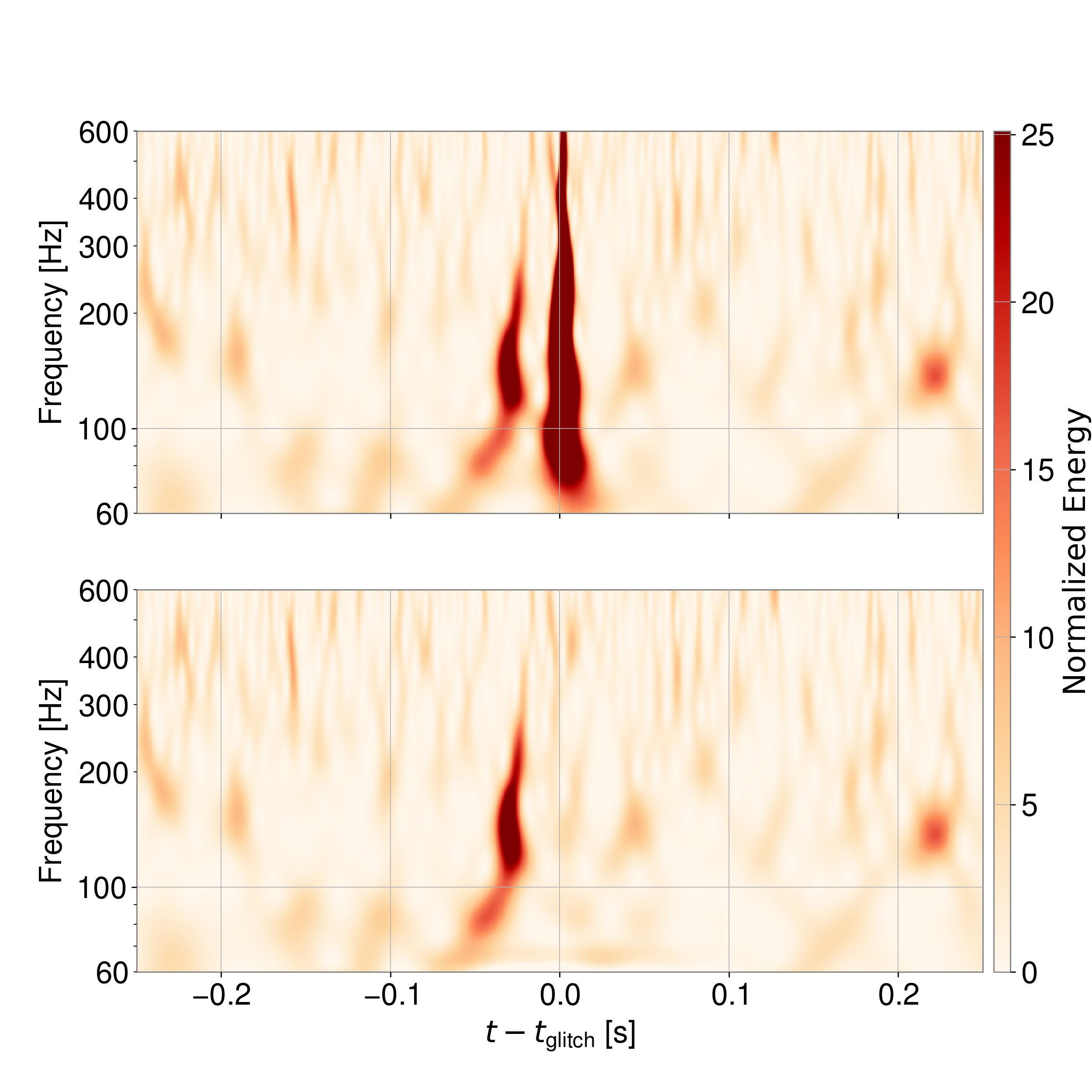}
    \caption{ Time-frequency representation of the \lho{} pre deglitched detector data containing blip glitch 1 and injected signal, $\bm{h} = \bm{s} + \bm{g} + \bm{n}$ (top panel), and the post deglitched detector data, $\bm{h}_{DG}=\bm{h}-\bm{g}_{rec}$ (bottom panel). The particular case shown here is one where the time of separation between the glitch and the injection is less than 0.005 seconds. }
    \label{fig:blip_cleaning_qscan}
\end{figure}

Figure \ref{fig:blip_cleaning_qscan} shows the time-frequency representation of \lho{} detector data pre and post deglitching for an example case of blip glitch 1. In the top panel, we see the blip glitch at $t - t_\mathrm{glitch} \sim 0$ seconds and the signal injection $0.005$ seconds before. The bottom panel shows the detector data with the glitch subtracted. By comparing it to the top panel, we see that \bw{} has subtracted the glitch from the data.

\begin{figure*}
    \includegraphics[width=1.2\columnwidth, clip=true]{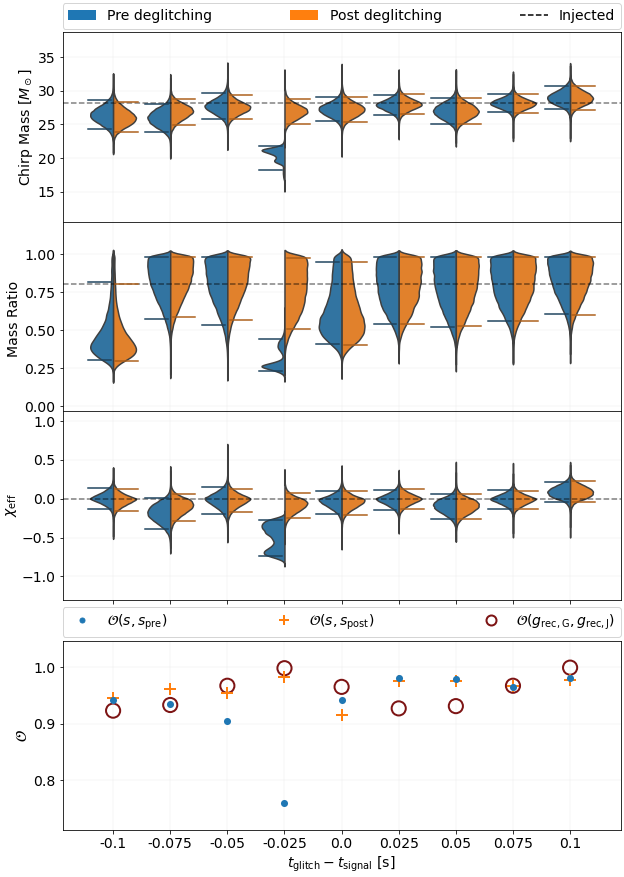}
    \caption{
    Analysis results for blip glitch 1. The top three panels plot the pre deglitching (blue) and post deglitching (orange) distributions for the chirp mass, the mass ratio and the effective spin ($\chi_{\mathrm{eff}}$) parameters as functions of the time separation between 
    the glitch ($t_{\mathrm{glitch}}$) and 
    signal injection ($t_{\mathrm{glitch}}$). The horizontal dashed lines show the injected value. The bottom panel shows the overlap between the maximum likelihood waveform by the \pe{} using the pre deglitched data (blue dot) and post deglitched data (orange plus). 
    }
    \label{fig:blip_pe_violin}
\end{figure*}

{To assess the impact of glitch subtraction on parameter estimation, we will compare posterior distributions produced using the pre- and post- deglitched for three different intrinsic parameters of the BBH system.  We will look at the chirp mass:} \begin{equation}
    \mathcal{M}_c=\frac{(m_1 m_2)^{3/5}}{(m_1+m_2)^{1/5}},
    \label{eq:chirpmasss}
\end{equation} 
{the mass ratio:}
\begin{equation}
    q=\frac{m_1}{m_2}
    \label{eq:massratio}
\end{equation} 
{where by convention $m_1\geq m_2$ such that $q\leq 1$, and the effective spin:}
\begin{equation}
    \chi_{\mathrm{eff}} = \frac{m_1\chi_1+m_2\chi_2}{m_1+m_2}.
    \label{eq:chieff}
\end{equation}
We present the results for these three quantities because they are better constrained than the individual masses and spins.

\begin{figure*}
    \includegraphics[width=1.2\columnwidth, clip=true]{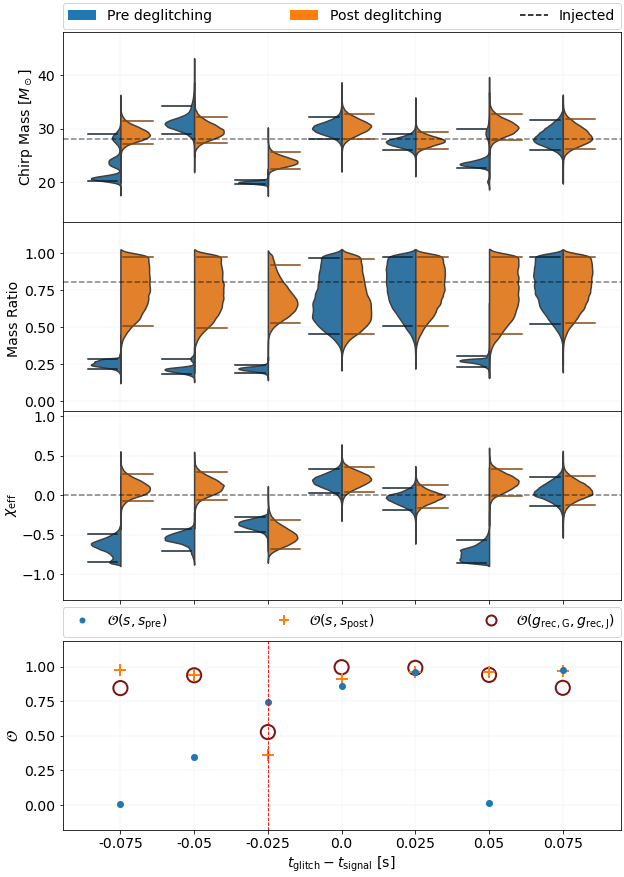}
    \caption{ Analog of Figure \ref{fig:blip_pe_violin} for blip glitch 2. Note that for $t_{\mathrm{glitch}} - t_{\mathrm{signal}} = -0.025$ seconds, both the pre-deglitching and post-deglitching posteriors exclude the injected values at the 90\% credible level.}
    \label{fig:blip2_pe_violin}
\end{figure*}

The top three panels of \ref{fig:blip_pe_violin}, \ref{fig:blip2_pe_violin}, and \ref{fig:blip3_pe_violin} show the posterior distributions of chirp mass, mass ratio, and  $\chi_{\mathrm{eff}}$ as functions of the separation time between the signal injection and the glitch. When the peak time of the glitch is more than 0.05 seconds away from the time of the coalescence, the posteriors from the pre-deglitched data are consistent with the true value, the glitch has minimal impact on PE as can be seen in Figure \ref{fig:blip_pe_violin}. This is expected, given the short time-duration of the blip. 
If the temporal separation between the glitch and time coalescence is less than 0.05 seconds, however, we see that the posteriors before deglitching are more likely to be biased from the true, injected values. The \pe{} posteriors on the deglitched data, however, are consistent with the true value. 

We see a similar trend in the overlap between the injected waveform and each of the maximum likelihood \pe{} waveforms of the pre deglitched (\preml) and post deglitched (\postml) posteriors (blue circle and orange plus respectively in panel 4 of Figures \ref{fig:blip_pe_violin}, \ref{fig:blip2_pe_violin}, and \ref{fig:blip3_pe_violin}). The overlaps between the injection and \preml{} are at lower values compared to those typically expected for the mass range of this injection set at times when the glitch and the signal are within 0.05 seconds \cite{Ghonge:2020suv}. The overlaps between the injection and \postml{} are at values consistent with the expected values from \cite{Ghonge:2020suv} indicating that the \postml{} is successfully able to recover the true waveform due to efficient deglitching. The panels also show the overlap between the median glitch reconstruction obtained from the \splusg{} model, henceforth called \gsplusg{} and the median glitch reconstruction obtained by \bw{}, henceforth called \ggo{} (red circle). We see that the overlaps are close to 1 indicating that \bw{} faithfully reconstructs the glitch waveform when the data contain an astrophysical signal just as efficiently as when there is no astrophysical signal. This means that \gsplusg{} does not contain any signal contamination.

\begin{table}
\clearpage{}\begin{tabular}{l|ll|ll|ll|}
\cline{2-7}
 & \multicolumn{2}{c|}{Chirp Mass \{[}\}$M_\odot$\{]\} & \multicolumn{2}{c|}{Mass Ratio} & \multicolumn{2}{c|}{$\chi_{eff}$} \\ \hline
\multicolumn{1}{|l|}{$t_{\rm glitch}-t_{\rm signal}$} & \multicolumn{1}{c|}{Before} & \multicolumn{1}{c|}{After} & \multicolumn{1}{c|}{Before} & \multicolumn{1}{c|}{After} & \multicolumn{1}{c|}{Before} & \multicolumn{1}{c|}{After} \\ \hline
\multicolumn{1}{|l|}{-0.075} & \multicolumn{1}{l|}{$\checkmark$} & $\checkmark$ & \multicolumn{1}{l|}{$\checkmark$} & $\checkmark$ & \multicolumn{1}{l|}{$\checkmark$} & $\checkmark$ \\ \hline
\multicolumn{1}{|l|}{-0.05} & \multicolumn{1}{l|}{} & $\checkmark$ & \multicolumn{1}{l|}{$\checkmark$} & $\checkmark$ & \multicolumn{1}{l|}{$\checkmark$} & $\checkmark$ \\ \hline
\multicolumn{1}{|l|}{-0.025} & \multicolumn{1}{l|}{$\checkmark$} & $\checkmark$ & \multicolumn{1}{l|}{} & $\checkmark$ & \multicolumn{1}{l|}{} & $\checkmark$ \\ \hline
\multicolumn{1}{|l|}{0.0} & \multicolumn{1}{l|}{} & $\checkmark$ & \multicolumn{1}{l|}{$\checkmark$} & $\checkmark$ & \multicolumn{1}{l|}{$\checkmark$} & $\checkmark$ \\ \hline
\multicolumn{1}{|l|}{0.025} & \multicolumn{1}{l|}{$\checkmark$} & $\checkmark$ & \multicolumn{1}{l|}{$\checkmark$} & $\checkmark$ & \multicolumn{1}{l|}{$\checkmark$} & $\checkmark$ \\ \hline
\multicolumn{1}{|l|}{0.05} & \multicolumn{1}{l|}{$\checkmark$} & $\checkmark$ & \multicolumn{1}{l|}{$\checkmark$} & $\checkmark$ & \multicolumn{1}{l|}{$\checkmark$} & $\checkmark$ \\ \hline
\multicolumn{1}{|l|}{0.075} & \multicolumn{1}{l|}{$\checkmark$} & $\checkmark$ & \multicolumn{1}{l|}{$\checkmark$} & $\checkmark$ & \multicolumn{1}{l|}{$\checkmark$} & $\checkmark$ \\ \hline
\end{tabular}
\clearpage{}
\caption{Table summarizing the PE preformance for blip glitch 1.  For each injection, the table notes whether the injected value was within the $90\%$ credible interval pre- and post-deglitching. A checkmark ($\checkmark$) means the injected value is within the $90\%$ CI, and an empty cell means it is not. }
\label{tab:Blip1}
\end{table}

We note, however, a poor deglitching performance in one 
case of blip glitch 2 where 
the signal is injected 0.025 s before the glitch, as shown in Figure \ref{fig:blip2_pe_violin}.
We see that both the pre deglitched and post deglitched posteriors exclude the injected values, indicating that deglitching the data did not have the bias mitigating effect we see for the corresponding case in blip glitch 1. We also observe a lower overlap of the injection with the \postml{} waveform, and a lower overlap between \gsplusg{} and \ggo{}. Our investigations of the \bw{} analysis show that the \splusg{} model does not correctly isolate the signal and glitch parts of the data (Figure \ref{fig:joint_model_reconstruction_blip2} in the appendix). The deglitched data suffers from a bias that exists due to uncleaned residual glitch power. This agrees with the findings from \cite{Hourihane:2022doe} for the same injection set. The proximity of the glitch with the signal merger causes a similar outlier in the quality of the signal reconstruction. However, the biasing effect in \cite{Hourihane:2022doe} is less severe compared to our case as the signal part of the model there is restricted to reconstructing a \cbc{} morphology which leads to a more effective reconstruction of the signal and glitch parts of the data.

\begin{table}
\clearpage{}\begin{tabular}{l|ll|ll|ll|}
\cline{2-7}
 & \multicolumn{2}{c|}{Chirp Mass {[}$M_\odot${]}} & \multicolumn{2}{c|}{Mass Ratio} & \multicolumn{2}{c|}{$\chi_{eff}$} \\ \cline{2-7} 
 & \multicolumn{1}{c|}{Before} & \multicolumn{1}{c|}{After} & \multicolumn{1}{c|}{Before} & \multicolumn{1}{c|}{After} & \multicolumn{1}{c|}{Before} & \multicolumn{1}{c|}{After} \\ \hline
\multicolumn{1}{|l|}{-0.075} & \multicolumn{1}{l|}{$\checkmark$} & $\checkmark$ & \multicolumn{1}{l|}{ } & $\checkmark$ & \multicolumn{1}{l|}{ } & $\checkmark$ \\ \hline
\multicolumn{1}{|l|}{-0.05} & \multicolumn{1}{l|}{ } & $\checkmark$ & \multicolumn{1}{l|}{ } & $\checkmark$ & \multicolumn{1}{l|}{ } & $\checkmark$ \\ \hline
\multicolumn{1}{|l|}{-0.025} & \multicolumn{1}{l|}{ } &   & \multicolumn{1}{l|}{ } & $\checkmark$ & \multicolumn{1}{l|}{ } &   \\ \hline
\multicolumn{1}{|l|}{0.0} & \multicolumn{1}{l|}{ } & $\checkmark$ & \multicolumn{1}{l|}{$\checkmark$} & $\checkmark$ & \multicolumn{1}{l|}{ } &   \\ \hline
\multicolumn{1}{|l|}{0.025} & \multicolumn{1}{l|}{$\checkmark$} & $\checkmark$ & \multicolumn{1}{l|}{$\checkmark$} & $\checkmark$ & \multicolumn{1}{l|}{$\checkmark$} & $\checkmark$ \\ \hline
\multicolumn{1}{|l|}{0.05} & \multicolumn{1}{l|}{$\checkmark$} & $\checkmark$ & \multicolumn{1}{l|}{ } & $\checkmark$ & \multicolumn{1}{l|}{ } & $\checkmark$ \\ \hline
\multicolumn{1}{|l|}{0.075} & \multicolumn{1}{l|}{$\checkmark$} & $\checkmark$ & \multicolumn{1}{l|}{$\checkmark$} & $\checkmark$ & \multicolumn{1}{l|}{$\checkmark$} & $\checkmark$ \\ \hline
\end{tabular}
\clearpage{}
\caption{Analog of Table~\ref{tab:Blip1} for blip glitch 2.}
\label{tab:Blip2}
\end{table}

Overall, our results show us that the deglitching procedure can mitigate the biasing caused by the blip glitches, and builds on the results of \cite{bayeswave_III} by showing the impact on PE. These results also agree with the results of \cite{Chatziioannou:2021ezd} and \cite{Hourihane:2022doe} which show successful glitch mitigation of the blip glitch class.

\subsection{Tomte} 

\begin{figure}
    \includegraphics[width=\columnwidth, clip=true]{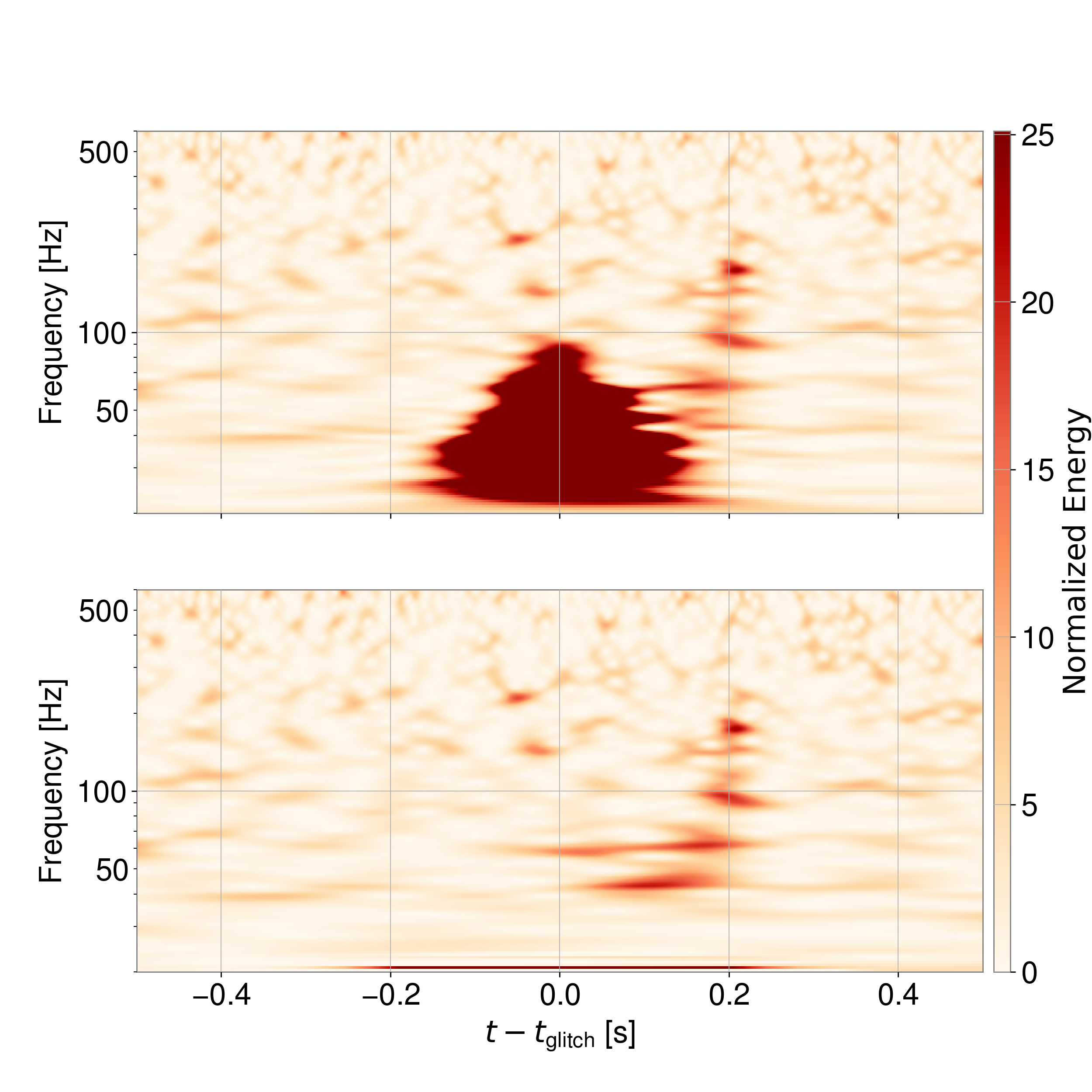}
    \caption{ Tomte glitch 1 analog of Figure \ref{fig:blip_cleaning_qscan}. Plot shows the pre and post deglitched data along with the injection in \llo{}. Note that since the glitch SNR is equal to 35 and the signal SNR in \llo{} is 12.4, the latter appears lighter compared to the signal in Figure \ref{fig:blip_cleaning_qscan}   }
    \label{fig:tomte_cleaning_qscan}
\end{figure}

\begin{figure*}
    \includegraphics[width=1.3\columnwidth, clip=true]{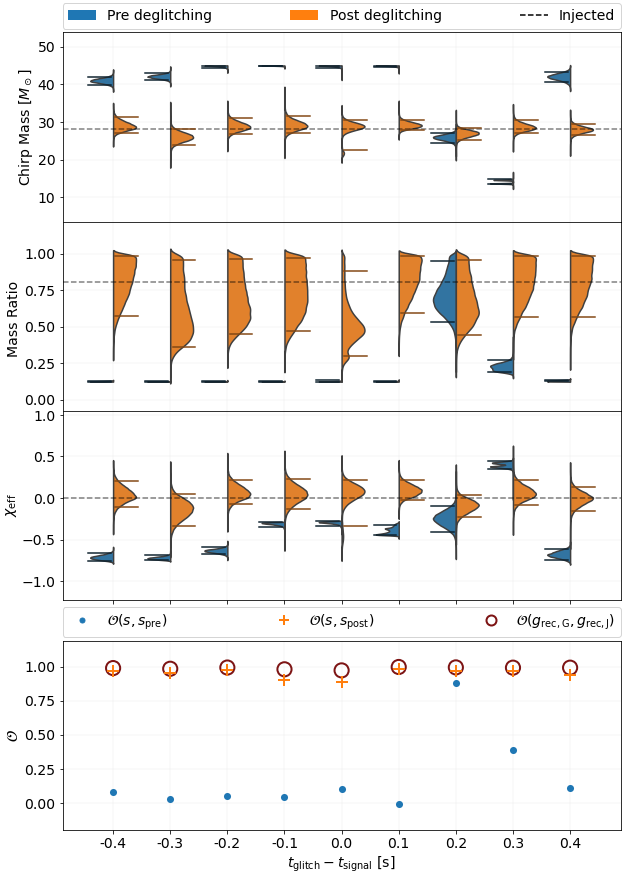}
    \caption{Analog of Figure \ref{fig:blip_pe_violin} for tomte glitch 1.}
    \label{fig:tomte_pe_violin}
\end{figure*}
We analyze two tomte glitches in \llo{} at GPS times 1243679045.58 (\textbf{4 June 2019}), and 1244425350.25 (\textbf{12 June 2019}) with \snr{s} 35 and 22.1 respectively as calculated by the Omicron pipeline \cite{Omicron}. We use the same configuration as for blip glitches, with an analysis segment of 4 seconds and a window of 1 second around the time of the signal.
We show the time-frequency representation of the pre and post deglitched detector data of tomte glitch 1 in Figure \ref{fig:tomte_cleaning_qscan}. The top panel shows the glitch power at $t-t_\mathrm{glitch} =$ 0 seconds and the signal power 0.2 seconds later. The bottom panel shows the data with the glitch power successfully removed.
We find that the presence of tomte glitches in the PE analysis window leads to posterior distributions that regularly exclude the true value at greater than 90\% probability in agreement with \cite{Hourihane:2022doe}. In fact, in almost all cases, the recovered posterior is sharply peaked at values far from the true value.
This is because tomte glitches peak at a frequency very close to the peak frequency of the binary black hole mergers we use in this study, 
and also have a smaller bandwidth compared to blip glitches. 
The post deglitching distributions include the injected value at the 90\% probability level and in cases of the chirp mass and $\chi_\mathrm{eff}$, the distribution peaks close to the true value ( Figures \ref{fig:tomte_pe_violin} and \ref{fig:tomte2_pe_violin}).  

\begin{table}
\clearpage{}\begin{tabular}{l|ll|ll|ll|}
\cline{2-7}
 & \multicolumn{2}{l|}{Chirp Mass {[}$M_\odot${]}} & \multicolumn{2}{l|}{Mass Ratio} & \multicolumn{2}{l|}{$\chi_{eff}$} \\ \hline
\multicolumn{1}{|l|}{$t_{\rm glitch}-t_{\rm signal}$} & \multicolumn{1}{l|}{Before} & After & \multicolumn{1}{l|}{Before} & After & \multicolumn{1}{l|}{Before} & After \\ \hline
\multicolumn{1}{|l|}{-0.4} & \multicolumn{1}{l|}{ } & $\checkmark$ & \multicolumn{1}{l|}{ } & $\checkmark$ & \multicolumn{1}{l|}{ } & $\checkmark$ \\ \hline
\multicolumn{1}{|l|}{-0.3} & \multicolumn{1}{l|}{ } & $\checkmark$ & \multicolumn{1}{l|}{ } & $\checkmark$ & \multicolumn{1}{l|}{ } & $\checkmark$ \\ \hline
\multicolumn{1}{|l|}{-0.2} & \multicolumn{1}{l|}{ } & $\checkmark$ & \multicolumn{1}{l|}{ } & $\checkmark$ & \multicolumn{1}{l|}{ } & $\checkmark$ \\ \hline
\multicolumn{1}{|l|}{-0.1} & \multicolumn{1}{l|}{ } & $\checkmark$ & \multicolumn{1}{l|}{ } & $\checkmark$ & \multicolumn{1}{l|}{ } & $\checkmark$ \\ \hline
\multicolumn{1}{|l|}{0.0} & \multicolumn{1}{l|}{ } & $\checkmark$ & \multicolumn{1}{l|}{ } & $\checkmark$ & \multicolumn{1}{l|}{ } & $\checkmark$ \\ \hline
\multicolumn{1}{|l|}{0.1} & \multicolumn{1}{l|}{ } & $\checkmark$ & \multicolumn{1}{l|}{ } & $\checkmark$ & \multicolumn{1}{l|}{ } & $\checkmark$ \\ \hline
\multicolumn{1}{|l|}{0.2} & \multicolumn{1}{l|}{ } & $\checkmark$ & \multicolumn{1}{l|}{$\checkmark$} & $\checkmark$ & \multicolumn{1}{l|}{ } & $\checkmark$ \\ \hline
\multicolumn{1}{|l|}{0.3} & \multicolumn{1}{l|}{ } & $\checkmark$ & \multicolumn{1}{l|}{ } & $\checkmark$ & \multicolumn{1}{l|}{ } & $\checkmark$ \\ \hline
\multicolumn{1}{|l|}{0.4} & \multicolumn{1}{l|}{ } & $\checkmark$ & \multicolumn{1}{l|}{ } & $\checkmark$ & \multicolumn{1}{l|}{ } & $\checkmark$ \\ \hline
\end{tabular}
\clearpage{}
\caption{Analog of Table~\ref{tab:Blip1} for Tomte glitch 1.}
\end{table}
We see a similar result from the overlap values in the bottom panels of Figures \ref{fig:tomte_pe_violin} and \ref{fig:tomte2_pe_violin} The overlaps between the \preml{} and the injection are close to 0 as the tomte glitch strongly biases \pe{}. The overlaps between \postml{} and the injection are close to 1 as is expected from \cite{Ghonge:2020suv}.  The overlaps between \gsplusg{} and \ggo{} are also close to 1, except in one case of tomte glitch 2 (glitch at 0.3 seconds ahead of the injection).

These results indicate that our \bw{} deglitching procedure can significantly mitigate the biasing impact of tomte glitches on \pe{}.

\subsection{Scattering}

\begin{figure}
    \includegraphics[width=\columnwidth, clip=true]{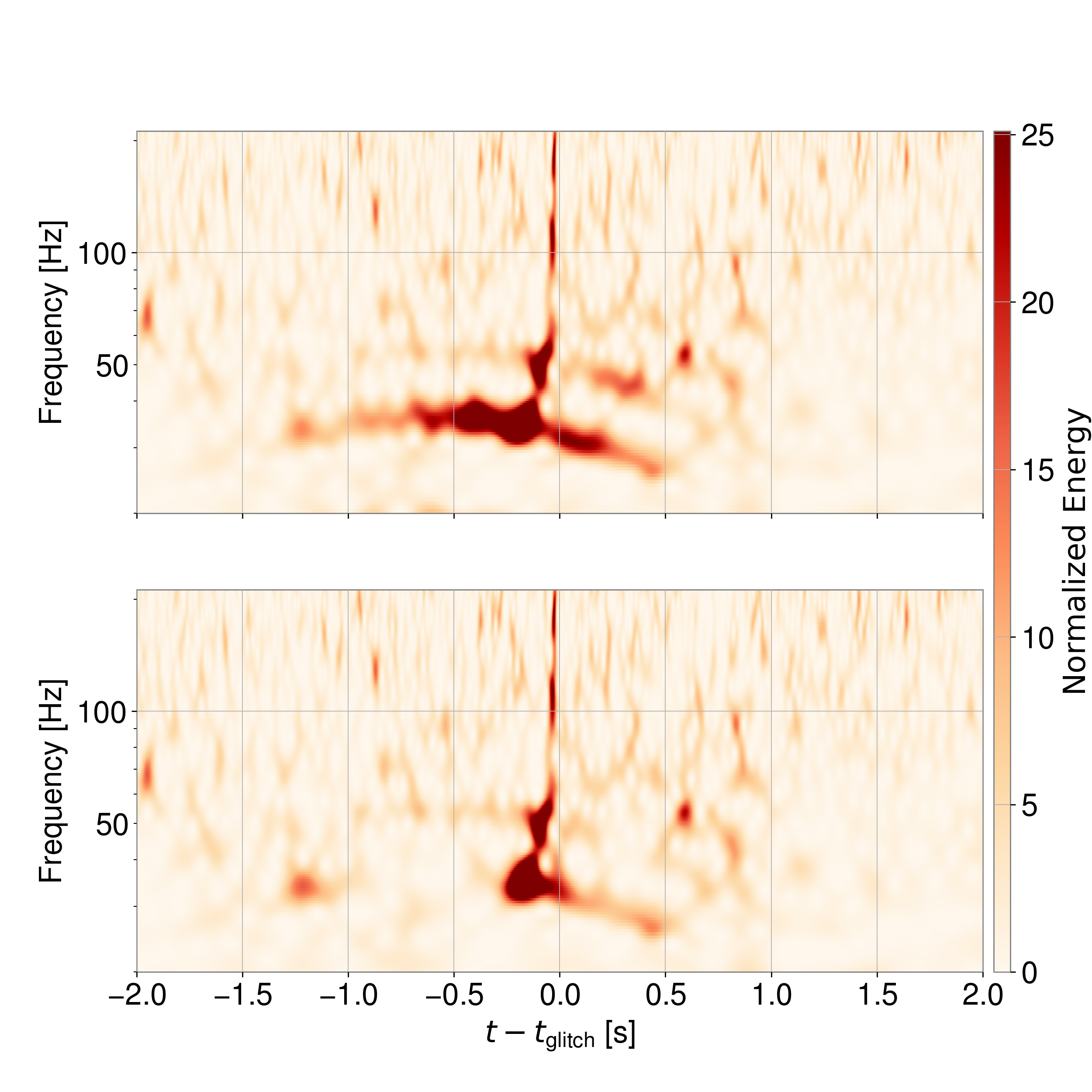}
    \caption{ Scattering glitch 1 analog of Figure \ref{fig:blip_cleaning_qscan} }
    \label{fig:scattering_cleaning_qscan}
\end{figure}

We perform a similar analysis with three scattering glitches at GPS times 1172917780 (\textbf{7 Mar. 2017}), 1166358283 (\textbf{21 Dec. 2016}), and 1177523957 (\textbf{29 Apr. 2017}) with \snr{s} 25, 13, and 15 respectively as calculated by the Omicron pipeline \cite{Omicron}. Due to the slightly longer duration of scattering glitches compared to blip and tomte glitches, the injections are separated by 0.2 seconds. For the \bw{} deglitching process, we use the modified priors introduced in Section \ref{subsec: priors}. 
 \begin{figure*}[h]
    \includegraphics[width=1.2\columnwidth, clip=true]{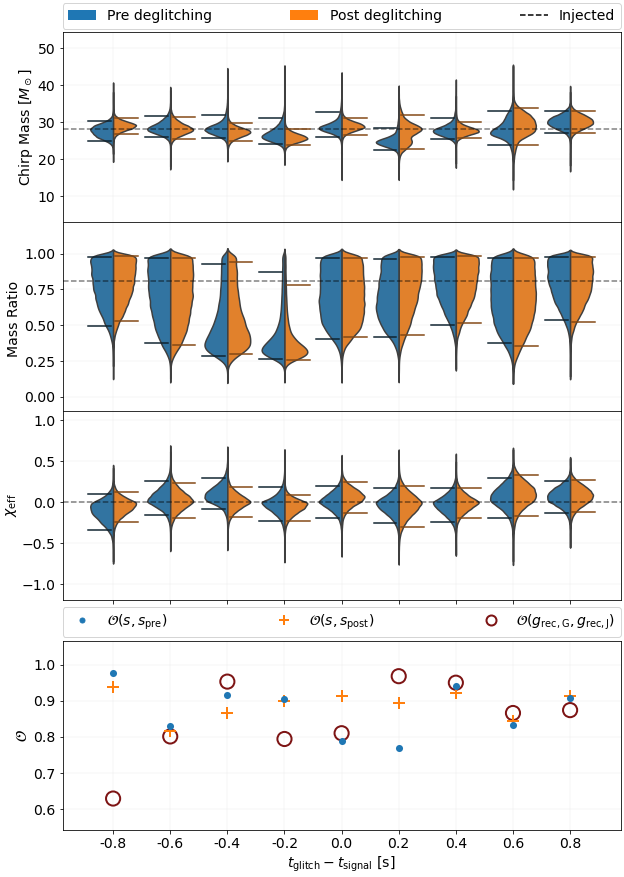}
    \caption{Analog of Figure \ref{fig:blip_pe_violin} for scattering glitch 1. }
    \label{fig:scattering_pe_violin}
\end{figure*}

\begin{table}
\clearpage{}\resizebox{0.9\columnwidth}{!}{
\begin{tabular}{l|ll|ll|ll|}
\cline{2-7}
 & \multicolumn{2}{l|}{Chirp Mass {[}$M_\odot${]}} & \multicolumn{2}{l|}{Mass Ratio} & \multicolumn{2}{l|}{$\chi_{eff}$} \\ \hline
\multicolumn{1}{|l|}{$t_{\rm glitch}-t_{\rm signal}$} & \multicolumn{1}{l|}{Before} & After & \multicolumn{1}{l|}{Before} & After & \multicolumn{1}{l|}{Before} & After \\ \hline
\multicolumn{1}{|l|}{-0.8} & \multicolumn{1}{l|}{$\checkmark$} & $\checkmark$ & \multicolumn{1}{l|}{$\checkmark$} & $\checkmark$ & \multicolumn{1}{l|}{$\checkmark$} & $\checkmark$ \\ \hline
\multicolumn{1}{|l|}{-0.6} & \multicolumn{1}{l|}{$\checkmark$} & $\checkmark$ & \multicolumn{1}{l|}{$\checkmark$} & $\checkmark$ & \multicolumn{1}{l|}{$\checkmark$} & $\checkmark$ \\ \hline
\multicolumn{1}{|l|}{-0.4} & \multicolumn{1}{l|}{$\checkmark$} & $\checkmark$ & \multicolumn{1}{l|}{$\checkmark$} & $\checkmark$ & \multicolumn{1}{l|}{$\checkmark$} & $\checkmark$ \\ \hline
\multicolumn{1}{|l|}{-0.2} & \multicolumn{1}{l|}{$\checkmark$} &   & \multicolumn{1}{l|}{$\checkmark$} &   & \multicolumn{1}{l|}{$\checkmark$} & $\checkmark$ \\ \hline
\multicolumn{1}{|l|}{0.0} & \multicolumn{1}{l|}{$\checkmark$} & $\checkmark$ & \multicolumn{1}{l|}{$\checkmark$} & $\checkmark$ & \multicolumn{1}{l|}{$\checkmark$} & $\checkmark$ \\ \hline
\multicolumn{1}{|l|}{0.2} & \multicolumn{1}{l|}{$\checkmark$} & $\checkmark$ & \multicolumn{1}{l|}{$\checkmark$} & $\checkmark$ & \multicolumn{1}{l|}{$\checkmark$} & $\checkmark$ \\ \hline
\multicolumn{1}{|l|}{0.4} & \multicolumn{1}{l|}{$\checkmark$} & $\checkmark$ & \multicolumn{1}{l|}{$\checkmark$} & $\checkmark$ & \multicolumn{1}{l|}{$\checkmark$} & $\checkmark$ \\ \hline
\multicolumn{1}{|l|}{0.6} & \multicolumn{1}{l|}{$\checkmark$} & $\checkmark$ & \multicolumn{1}{l|}{$\checkmark$} & $\checkmark$ & \multicolumn{1}{l|}{$\checkmark$} & $\checkmark$ \\ \hline
\multicolumn{1}{|l|}{0.8} & \multicolumn{1}{l|}{$\checkmark$} & $\checkmark$ & \multicolumn{1}{l|}{$\checkmark$} & $\checkmark$ & \multicolumn{1}{l|}{$\checkmark$} & $\checkmark$ \\ \hline
\end{tabular}
}
\clearpage{}
\caption{Analog of Table~\ref{tab:Blip1} for scattering glitch 1.}
\label{tab:Scattering1}
\end{table}
Since scattering glitches are typically longer ($>$2 seconds) than simulated \cbc{} signals used in this study
we configure the glitch part of the \splusg{} model to use a 2-second window, and the signal model to use a 1-second window.  We let the dimensionality of the glitch model increase to a maximum of 100 wavelets while limiting the signal model to 10 wavelets. 
By limiting the signal dimensionality to a maximum of 10, we preemptively exclude the signal model from reconstructing the glitch power. We also increase the maximum quality factor to 160 for the glitch model while limiting that of the signal model to 40 (See Sec.~\ref{subsec: priors} for a discussion on why we increase the maximum $Q$ for longer-duration glitches). We use a longer analysis segment length of 8 seconds to capture the full power content of scattering glitches given their characteristic durations of 3–4 seconds, compared to the standard \bw{} default of 4 seconds used for blip and tomte glitches given their millisecond durations.

Figure \ref{fig:scattering_cleaning_qscan} shows the time-frequency representations of the pre and post deglitched data of an example injection near scattering glitch 1. Similar to Figures \ref{fig:blip_cleaning_qscan} and \ref{fig:tomte_cleaning_qscan} we see that deglitching removes the glitch power from the data while keeping the signal power intact.
 
Figures \ref{fig:scattering_pe_violin}, \ref{fig:scattering2_pe_violin}, and \ref{fig:scattering3_pe_violin} show scattering glitch has a minimal effect on \pe{} of chirp mass, mass ratio, and $\chi_\mathrm{eff}$ for our simulated binary black hole events.
The posteriors are consistent with the true values and similar in distribution to the posteriors after deglitching. 

The overlaps in the bottom panels of Figures \ref{fig:scattering_pe_violin},  \ref{fig:scattering2_pe_violin}, and \ref{fig:scattering3_pe_violin} also demonstrate that \preml{} and \postml{} waveforms agree with the injected waveform with overlap values expected from \cite{Ghonge:2020suv}. The overlap between \gsplusg{} and \ggo{} is also high for times when the separation between the time of the coalescence of the signal and the center of the glitch is small ($<$ 0.5 seconds). At times far away, the analysis window is unable to fully include the entire duration of the glitch, and we see a drop in the overlap. 

\bw{} can separate signal and glitch power due to the very different morphology of scattering glitches and stellar mass \bbh{} coalescences, as in the case of our injection with total mass $\sim 65 M_{\odot}$.
We find similar results when we inject a lower mass \bbh{} signal
($M_T \sim 19 M_{\odot}$) 
as shown in Figure \ref{fig:scattering_gw170608_violin}. The injected signal here has source parameters similar to \cite{gw170608_2017} and its luminosity distance is scaled such that its \snr{} is equal to 15.  This is in agreement with the results shown in \cite{Chatziioannou:2021ezd} and \cite{Hourihane:2022doe} which show that scattering glitches have a less severe impact on the \pe{} of stellar mass \bbh{} binaries compared to blip and tomte glitches. We caution the reader these results apply specifically to \bbh{} systems from $19 M_{\odot}$ to $65 M_{\odot}$ with signal SNR=15 and glitch SNRs as listed above. Further investigations on the relative impacts of the signal and glitch SNRs on the performance of the deglitching procedure will need to be performed in a future study.

 \section{Conclusion}
\label{sec:conclusion}

Transient noise artifacts, colloquially known as {\em glitches}, have persisted in the gravitational wave strain data from the LIGO-Virgo-KAGRA network. These artifacts can contaminate the signals from astrophysical sources, degrading the accuracy of parameter estimation. In this paper, we assess the efficacy of a procedure to subtract glitches from the signal data and improve the characterization of source parameters through Bayesian inference. The procedure involves modeling the signal, glitch, and noise parts of the data simultaneously using the \bw{} \splusg{} configuration introduced in \cite{bayeswave_III}. We apply the procedure to an SNR=15 injection with total mass $M_T = 65 M_{\odot}$ with properties similar to GW150914 \cite{PEofGW150914}. The conclusions present here may not necessarily apply to qualitatively different systems such as \nsbh{} and \bns{} systems where the interplay between the signal and glitch parts of the data will need to be more finely tuned for effective deglitching using \bw{.} Additionally, most of the glitch instances presented here have SNRs greater than the signal. For other combinations of signal and glitch SNRs, the impact of the glitch on PE and the role of deglitching will need to be studied. The modified prior ranges presented in this work may be helpful in this fine-tuning. \cite{Hourihane:2022doe} provides more detailed studies in this regard. 

We find that blip glitches moderately affect parameter estimation results when the glitch and CBC merger are coincident within $< 0.1$ seconds. {In cases where \bw{} can successfully model and subtract the glitch the accuracy of the parameter estimation is significantly improved,}
and the overlap between the injected waveform and recovered maximum likelihood waveform is $>0.9$, consistent with the values from \cite{Ghonge:2020suv}. 
{However, we note that there are instances in which \bw{} does not successfully subtract the glitch, such as the injection $0.025\ {\rm s}$ before blip glitch 2 (See. Fig.{\ref{fig:blip2_pe_violin}}). A primary cause of these failures is the signal model reconstructing part of the glitch power and vice versa. A more detailed look into one of these unsuccessful glitch models is included in Appendix {\ref{sec:AppA}}.}

Tomte glitches bear a strong resemblance to blips, but with their longer duration, pose a greater hindrance to parameter estimation. This is also reflected in the overlaps between the injected waveform and the recovered maximum likelihood waveform which are $<0.5$ in the majority of the cases. However, this study suggests that glitch subtraction is effective in reducing their impact and yielding accurate parameter estimates. The overlaps also increase to their nominal values of $>0.9$, consistent with \cite{Ghonge:2020suv}.

Finally, scattering glitches, with their long-duration narrow-band harmonics resembling the low-frequency inspiral phase of BBH mergers, do not seem to affect the results of parameter estimation, and the overlaps between the injected waveform and the recovered maximum likelihood waveform are $>0.9$ in the majority of the cases for both pre and post deglitched data. Possible future works include studying the impacts of these glitches on longer-duration signals such as 
\bns{} and \nsbh{} mergers using further fine-tuning to the individual model wavelet prior bounds which we introduce here.

A deglitching procedure similar to the one presented in this paper has been applied to past \gw{} detections by LIGO, Virgo, and KAGRA \cite{GWTC1, GWTC2, GWTC3}. This paper provides a baseline on the BayesWave \splusg{} model's ability to mitigate some of the common glitch types encountered in LIGO's first three observing runs, which are likely to persist in these future observations. 
{As also emphasized in works such as} \cite{Chatziioannou:2021ezd}, \cite{Hourihane:2022doe}, and \cite{Davis:2022dnd} {the deglitching procedure is not an exact science.    All events that required deglitching in the first three \gw{} transient catalogs have undergone thorough reviews, including  the residuals test described in} \cite{Davis:2022dnd}.  {These tests are not fully automated and require visual inspection and human intervention.  Furthermore, characteristics of detector noise changes between observing runs, and even within observing runs, therefore glitch subtraction procedures and reviews should be re-evaluated regularly.}

The modified \bw{} model described in \cite{Chatziioannou:2021ezd} provides a more accurate reconstruction of the \cbc{} signal and allows a more distinct resolution between the signal and glitch components. Our work is complementary to \cite{Chatziioannou:2021ezd} and \cite{Hourihane:2022doe} as it demonstrates that deglitching is effective without a known signal model, and can be used in cases when modeling the signal as a \cbc{} may not be as effective such as in the cases of core-collapse supernovae \cite{Szczepanczyk:2021bka} or post-merger \bns{} signals \cite{Chatziioannou:2017ixj} overlapping with glitches. It can also be used as a consistency check for the deglitching process if the astrophysical signal is suspected to have higher-order spherical harmonic modes (HOM) \cite{Khan:2019kot}. We point to possible future work including integrating the glitch priors developed in this work with the ones described in \cite{Hourihane:2022doe}, and automating the deglitching process by using event trigger generators, such as the Omicron pipeline \cite{Omicron}, to configure the \bw{} pipeline with an event-specific set of priors to achieve more efficient deglitching. 

 \section{Acknowledgments}
\label{acknowledgments}

This research has made use of data, software and/or
web tools obtained from the Gravitational Wave Open
Science Center (https://www.gw-openscience.org), a service of LIGO Laboratory, the LIGO Scientific Collaboration and the Virgo Collaboration. LIGO is funded by the U.S. National Science Foundation. Virgo is funded by the French Centre National de Recherche Scientifique (CNRS), the Italian Istituto Nazionale della Fisica Nucleare (INFN) and the Dutch Nikhef, with contributions by Polish and Hungarian institutes. The authors are grateful for computational resources provided by the LIGO Laboratory and supported by National Science Foundation Grants PHY-0757058 and PHY-0823459.

 This research was done using services provided by the OSG Consortium \cite{osg07,osg09, https://doi.org/10.21231/906p-4d78, https://doi.org/10.21231/0kvz-ve57}, which is supported by the National Science Foundation awards \#2030508 and \#1836650. 
 
 The GT authors gratefully acknowledge the NSF for support from Grants PHY-1809572 and PHY-2110481. KC and SH were supported by NSF Grant PHY-2110111. SH is supported by the National Science Foundation Graduate Research Fellowship under Grant No. DGE-1745301. \appendix
\section{Additional Details on an Unsuccessful Subtraction}
\label{sec:AppA}
As stated in Sec {\ref{sec:conclusion}}, glitch subtraction for events in the gravitational wave transient catalogs \cite{GWTC1,GWTC2,GWTC3} undergo internal review.  Since detector noise (transient and stationary) can be unpredictable and not all \gw{} events are the same, each glitch subtraction case may be different.
Here we look more in depth at one  instance where \bw{} deglitching is unsuccessful, and how such a case may be identified even if the true signal is unknown.

\begin{figure}[ht]
    \includegraphics[width=0.9\columnwidth,clip=true]{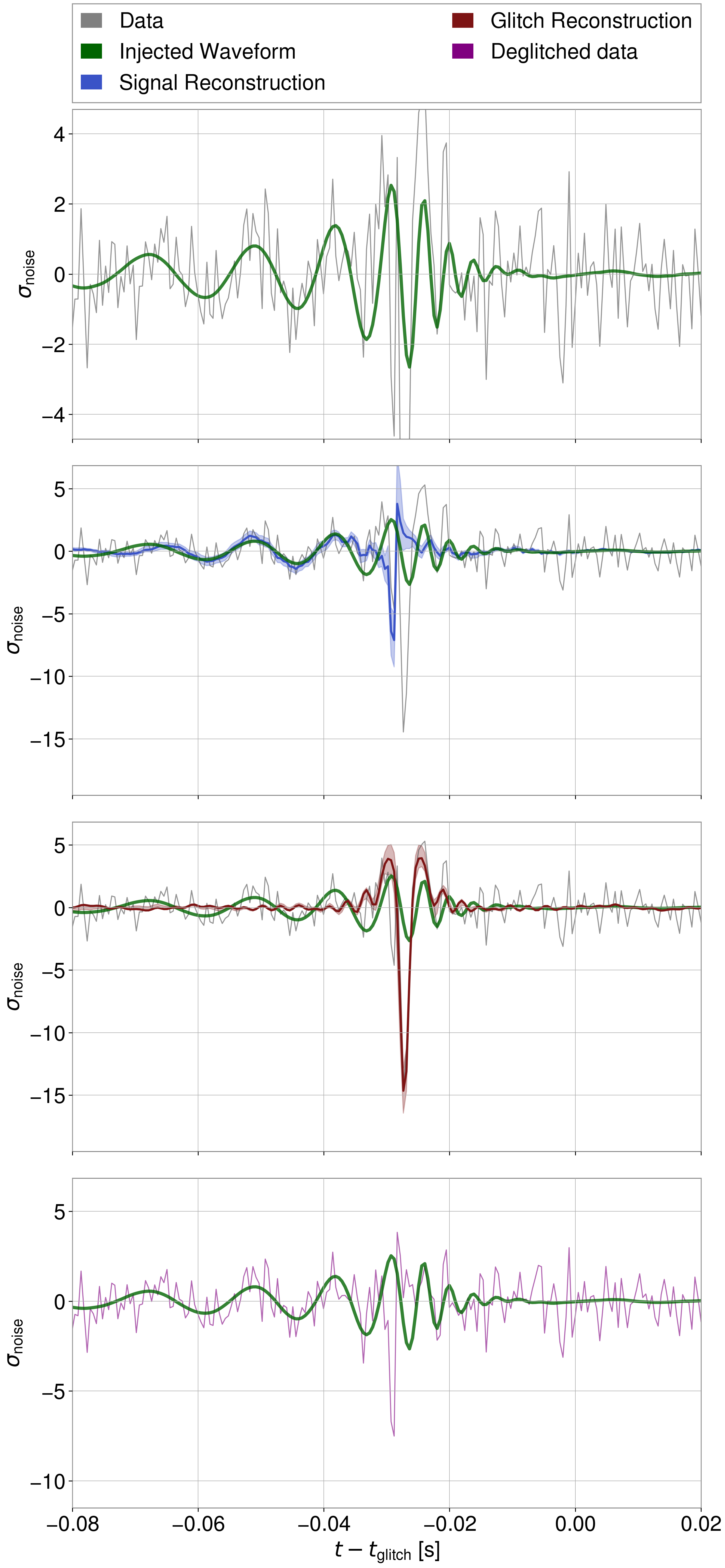}
    \caption{Analog of Figure \ref{fig:joint_model_reconstruction} for 
    blip glitch 2}
    \label{fig:joint_model_reconstruction_blip2}
\end{figure}

The event we highlight is the injection $0.025\ {\rm s}$ after blip glitch 2.  Figure~\ref{fig:blip2_pe_violin} shows that the posterior distributions for the chirp mass and $\chi_{eff}$ are still narrowly peaked away from the true value even after glitch subtraction.  The bottom panel of Figure~\ref{fig:blip2_pe_violin} also shows the overlap between the injected waveform and the maximum-likelihood waveform from parameter estimation are below 0.75, much lower than the overlaps of $\geq.9$ we see for other injections (and in the glitch-subtracted data has a lower overlap than the original data).
Of course for a real \gw{} signal we will not know the true waveform, and so we need to look at the subtracted data and \bw{} reconstruction of the signal and glitch models. Figure {\ref{fig:joint_model_reconstruction_blip2}} shows from top to bottom: the original data, the \bw{} signal reconstruction, the \bw{} glitch reconstruction, and the deglitched data. The injected signal is shown in green in all four panels for reference.  From visual inspection, we can see that the \bw{} signal model does not look like a standard BBH signal, and that it is picking up some of the blip glitch power. The bottom panel shows that a significant amount of power remains after glitch subtraction, with residuals around $-0.01\ {\rm s}$ at more than $5\sigma$ deviation.
This case would be flagged for further review even in the absence of a true known waveform.

\section{Additional Figures}
This appendix collects some additional results, with  violin and overlap plots similar to Figure \ref{fig:blip_pe_violin}  for glitch instances that are not included in the main body of the paper.

\begin{figure}[ht]
    \includegraphics[width=0.9\columnwidth,clip=true]{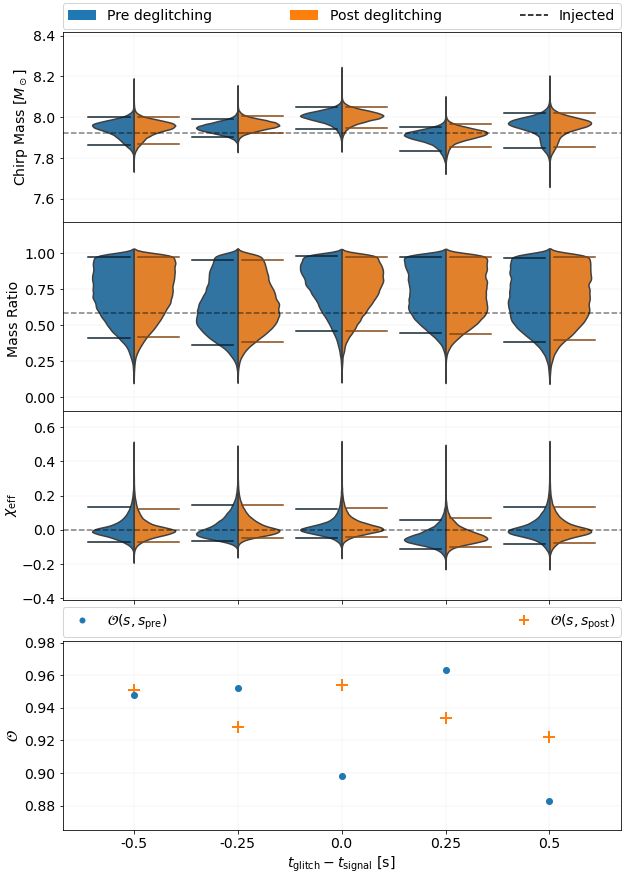}
    \caption{Analog of Figure \ref{fig:scattering_pe_violin} for 
    \bbh{} injections with $M_T \sim 19 M_{\odot}$ near scattering glitch 2}
    \label{fig:scattering_gw170608_violin}
\end{figure}

\begin{figure}[ht]
    \includegraphics[width=0.9\columnwidth, clip=true]{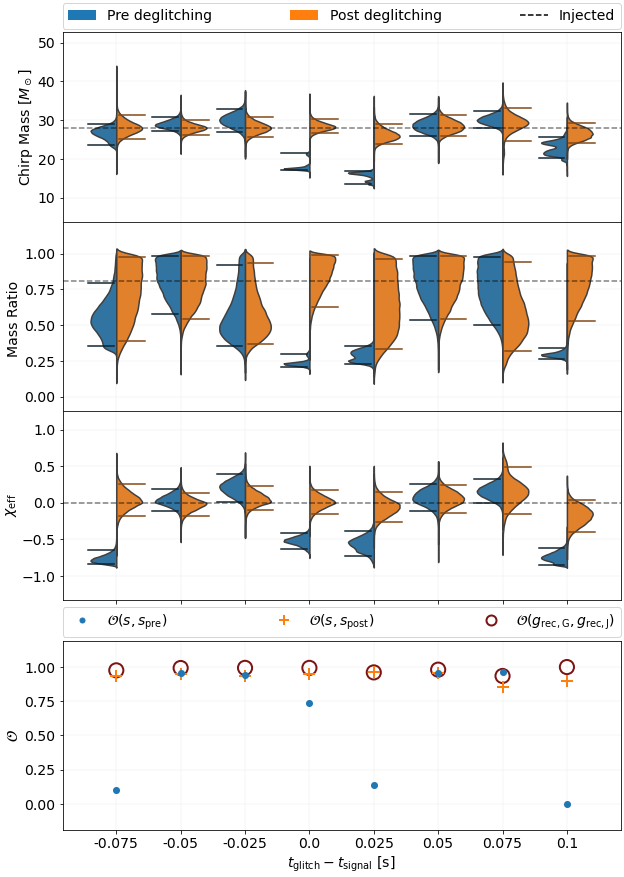}
    \caption{ Analog of Figure \ref{fig:blip_pe_violin} for blip glitch 3.}
    \label{fig:blip3_pe_violin}
\end{figure}

\begin{figure}[ht]
    \includegraphics[width=1\columnwidth, clip=true]{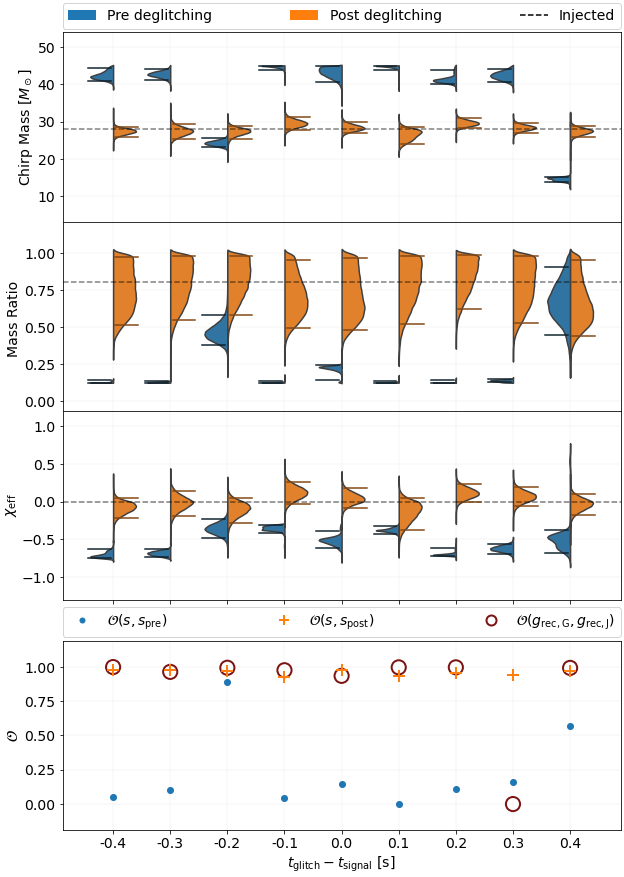}
    \caption{ Analog of Figure \ref{fig:tomte_pe_violin} for tomte glitch 2.}
    \label{fig:tomte2_pe_violin}
\end{figure}

\begin{figure}[ht]
    \includegraphics[width=1\columnwidth, clip=true]{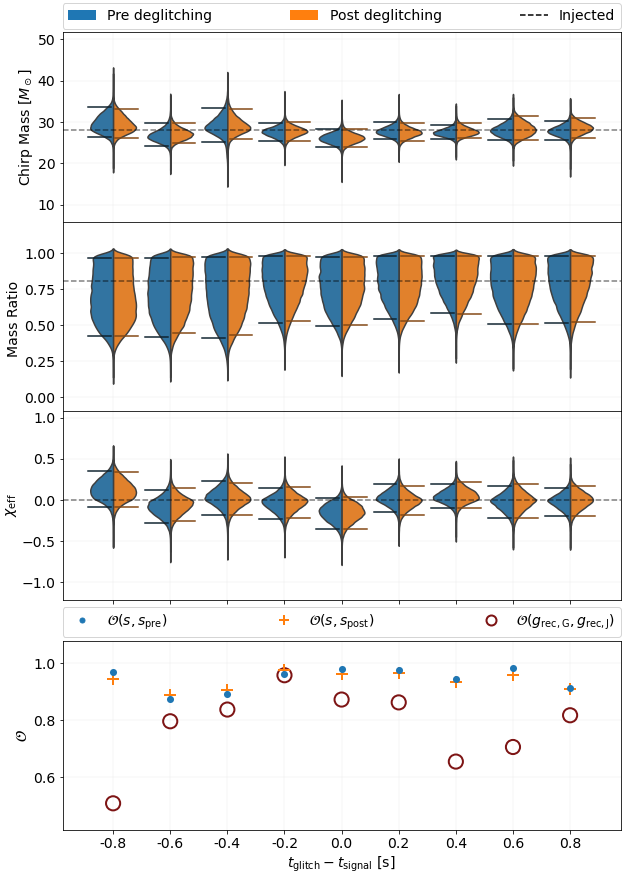}
    \caption{ Analog of Figure \ref{fig:scattering_pe_violin} for scattering glitch 2.}
    \label{fig:scattering2_pe_violin}
\end{figure}

\begin{figure}[ht]
    \includegraphics[width=1\columnwidth, clip=true]{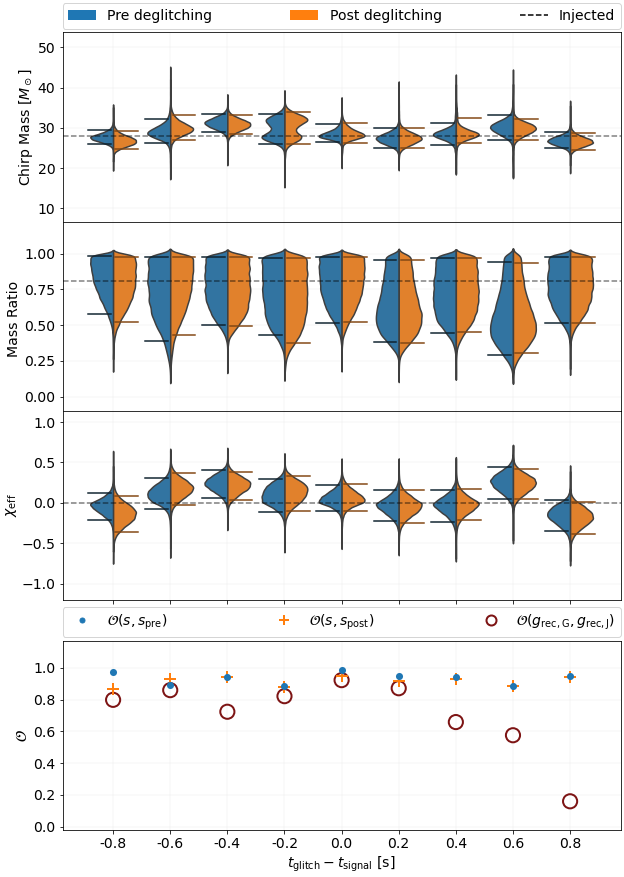}
    \caption{ Analog of Figure \ref{fig:scattering_pe_violin} for scattering glitch 3.}
    \label{fig:scattering3_pe_violin}
\end{figure} 

\clearpage
\bibliography{paper}
\end{document}